\documentclass[reprint,aps,pra,amsmath,amssymb,twocolumn,superscriptaddress]{revtex4-2}

\usepackage[english]{babel}
\usepackage{hyperref}
\usepackage{amsfonts}
\usepackage{multirow}
\usepackage{soul}

\usepackage[pdftex]{graphicx}
\usepackage{color}
\usepackage[toc,page]{appendix}
\usepackage{float}

\usepackage{amsmath, bbold}
\usepackage{amssymb}
\usepackage{mathtools}
\usepackage{braket}
\usepackage{cancel}

\usepackage{filecontents}
\usepackage[sort&compress]{natbib}

 \newcommand{\latdev}[1]{{\fontfamily{qcr}\selectfont #1}}

\begin{document}

\title{Towards a benchmark for quantum computers based on an iterated post-selective protocol}

\author{Adrian Ortega}
\email{ortega.adrian@wigner.hu}
\affiliation{HUN-REN Wigner RCP, Konkoly-Thege M. \'{u}t 29-33., H-1121 Budapest, Hungary}
\affiliation{Departamento de F\'isica, Universidad de Guadalajara, Blvd. Gral. Marcelino Garc\'ia Barrag\'an 1421, C.P. 44430, Guadalajara, Jalisco, Mexico}

\author{Orsolya K\'alm\'an}
\email{kalman.orsolya@wigner.hu}
\affiliation{HUN-REN Wigner RCP, Konkoly-Thege M. \'{u}t 29-33., H-1121 Budapest, Hungary}

\author{Tam\'as Kiss}
\email{kiss.tamas@wigner.hu}
\affiliation{HUN-REN Wigner RCP, Konkoly-Thege M. \'{u}t 29-33., H-1121 Budapest, Hungary}






\begin{abstract}
Applying post selection in each step of an iterated protocol leads to sensitive quantum dynamics that may be utilized to test and benchmark current quantum computers. An example of this type of protocols was originally proposed for the task of matching an unknown quantum state to a reference state. We propose to employ the quantum state matching protocol for the purpose of testing and benchmarking quantum computers. In particular, we implement this scheme on freely available IBM superconducting quantum computers. By comparing measured values with the theoretical conditional probability of the single, final post-selected qubit, which is easy to calculate classically, we define a  benchmark metric. Additionally, the standard deviation of the experimental results from their average serves as a secondary benchmark metric, characterizing fluctuations in the given device. A peculiar feature of the considered protocol is that there is a phase parameter of the initially prepared state, on which the resulting conditional probability should not depend. A careful analysis of the measured values indicates that its dependence on the initial phase can reveal useful information about coherent gate errors of the quantum device.

\end{abstract}

\maketitle

\section{Introduction}
\label{sec:Introduction}


In the field of classical computing, ``benchmarks are standard tools that allow evaluating and comparing different systems or components according to specific characteristics such as performance, dependability, and security''~\cite{Vieiraext2013}. There are variations to this definition (see e.g.~\cite{Liljabook2000}, and also \cite{Kounevbook2020} and references therein), but the one given here is perhaps the most exemplary for designing benchmarks for quantum computers as well. Although in classical computation performance, dependability and security are considered to be equally important, in current NISQ (Noisy Intermediate-Scale Quantum) computation, where errors are still significant and may hinder useful computation, we are still focusing mostly on the assessment of performance~\cite{ReschACM2021,ProctorNatPhys2021}. 



A quantum computer benchmark is understood, in general, as a task or a set of tasks that needs to be performed by the given device, and a subsequent assessment of this performance by the evaluation of a benchmark metric or metrics ~\cite{BlumeOSTI2019, AcuavivaArxiv2024,ProctorArxiv2024}). The set of benchmark tasks or tests run in NISQ systems aim either to assess certain components, such as gates, as is done in randomized benchmarking~\cite{KnillPRA2008,MagesanPRL2011,MagesanPRA2012,ProctorPRL2019} or apply a holistic approach and test all components simultaneously. There are two approaches in this latter direction: (1) applying randomized circuits, like in the Quantum Volume (QV) benchmark~\cite{CrossPRA2019}, the mirror cicuit benchmark~\cite{ProctorNatPhys2021,ProctorPRL2022} or Clifford-circuit benchmarking~\cite{ChenPRA2023}, or (2) implement a meaningful task, such as optimization~\cite{MesmanArxiv2021,LubinskiArxiv2023}, quantum information protocols ~\cite{LinkePNAS2017,WrightNatCom2019,GilyenArxiv2021}, as well as applications in quantum chemistry\cite{McCaskeyNQI2019}. Benchmark suites, involving the implementation of many basic applications used as benchmarks have are already been published~\cite{FinzgarArxiv2022,TomeshArxiv2022,LiArxiv2020}.  


We also have to mention that the so-called performance benchmarks in current NISQ devices fall in the category of "research benchmarks" geared towards quality results (e.g., does the executed quantum circuit provide a meaningful result and not just random data?); in contradistinction,  from the classical point of view~\cite{Kounevbook2020}, a classical performance benchmark usually means the amount of work done against time and resources needed. Up to now there are few investigations on ``time variable performance'' in NISQ systems, e.g.~\cite{DasguptaArxiv2020}, mainly because current quantum computers are unstable, in the sense that they have to be calibrated regularly, the chances of suffering decoherence are high, and quantum error correction is still not available to help to maintain the reliability in computations. Since nowadays NISQ systems are small and noisy, it seems important to emphasize that a meaningful benchmark metric in our opinion should be able to reveal the degree of the decoherence and inherent errors (coherent errors, measurement errors, etc.) in the tested devices to a certain degree. It is by no means clear how to design such a benchmark task that can address these issues~\cite{ProctorArxiv2024}.

Iterative, post-selection based protocols lead to an essentially nonlinear quantum dynamics, which may be highly sensitive to the initial conditions~\cite{KissPRA2006,GilyenSciRep2016} as well as noise and errors~\cite{MalachovChaos2019,ViennotCSF2022,GuanPRA2013,PortikPLA2022,PortikPRA2024}. Protocols with fully chaotic time evolution and ones realizing the Mandelbrot set (also characteristic of chaotic dynamics) have been proposed for benchmarking current quantum computers \cite{GilyenArxiv2021}. In addition to chaotic behavior, another interesting feature of iterated nonlinear protocols is that superattractive fixed points or cycles may be present in the dynamics, which guarantee that for a certain range of input parameters the behavior is ideally uniform and stable~\cite{TorresPRA2017}. 

In this paper, we propose to use a specific nonlinear protocol that is not only superattractive but adaptive as well so that the test procedure can be tailored to a broad range of possible cases. The protocol of quantum state matching was originally proposed for the task of deciding whether an unknown qubit state lies in the vicinity of a reference state or not \cite{KalmanPRA2018}. In Ref.~\cite{OrtegaPS2023} we have implemented a single step of this protocol in superconducting quantum computers with many different settings. Here we perform a thorough analysis of a specific quantity (the overall success probability after $n=1,2$ iterations) and define related benchmark metrics to assess the performance of the tested freely available IBM quantum computers. Based on our tests, we model possible sources of errors (incoherent as well as coherent gate errors) leading to deviations from the ideal theoretical results.

The paper is organized as follows. In Sec.~\ref{sec:implprot} we briefly describe the protocol that we use for testing quantum computers. In Sec.~\ref{sec:statframe} we explain the quantity that we intend to estimate from the measurements and the statistical considerations relevant for our studies. In Section~\ref{sec:res} we present the results of the various tests run on IBM quantum computers. In Section~\ref{sec:benchmark} we propose benchmark metrics and assess the performance of the tested devices. In Section~\ref{sec:heur} the test results are analyzed in detail in order to estimate error parameters of the tested devices. The last section contains our summary and the outlook.

\section{The test protocol}
\label{sec:implprot}
In the following, we describe briefly the quantum state matching scheme~\cite{KalmanPRA2018} that we will use as a test protocol to assess the performance of quantum computers. 

The quantum state matching protocol~\cite{KalmanPRA2018} was originally designed for the task of deciding whether an unknown qubit state $\ket{\Phi_{0}}$ (e.g. coming from a quantum computation, or state preparation) lies in a prescribed $\varepsilon$-neighborhood ($0<\epsilon\leq 1$) of a reference (desired) $\ket{\psi_{\rm ref}}$ quantum state. The protocol is applied iteratively on pairs of input states from an ensemble, and in every iteration, the same two-qubit unitary operation $U$ acts on the pairs of qubits that have been transformed by the previous iteration. A single iterational step is probabilistic, since after the action of the two-qubit unitary we measure one of the qubits and we only keep the other (unmeasured) one, if the measurement resulted '0'. The unitary itself is determined by the value of $\varepsilon$ and the amplitudes of the reference state. If the input (unknown) state lies inside (outside) the $\varepsilon$-neighborhood of (i.e., a circle of radius $\varepsilon$ around) $\ket{\psi_{\rm ref}}$, then with every successful iteration, the protocol will gradually take the unknown state closer and closer to the $\ket{\psi_{\rm ref}}$ reference state (to the state $\ket{\psi_{\rm ref}^{\perp}}$ orthogonal to the reference state) already in a few steps. 

\begin{figure}[htb]
  \centering
  \includegraphics[scale=0.5]{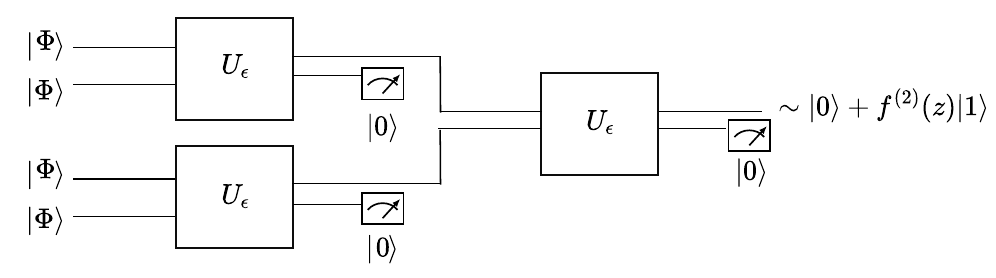}
  \caption{Quantum circuit for implementing $n$ iterations of the protocol. $f^{(2)}(z)$ denotes the function composition
  of the complex map $f(z)$ of Eq.~(\ref{eq:Phi1}). The circuit can be easily generalized for $n$ iterations of the protocol, for which $2^{n}$ qubits are necessary. Note that the intermediate measurements (and thus the post-selection) can be postponed to the end of the circuit.}
  \label{fig:circ_full}
\end{figure}

Mathematically, if we parameterize the input pure state as 
\begin{equation}
|\Phi_{0}\rangle=\mathcal{N}_{0} \left(|0\rangle + z|1\rangle\right),  
\label{eq:Phi0}
\end{equation}
where $z\in\mathbb{C}\cup\infty$ and $\mathcal{N}_0$ is a normalization factor, then the resulting quantum dynamics can be described by an iterated complex quadratic rational map $f(z)$, which transforms after $n$ iterations the input state into 
\begin{equation}
    |\Phi_n \rangle = \mathcal{N}_{n}\left(|0\rangle + f^{(n)}(z)|1\rangle\right),
    \label{eq:Phi1}
\end{equation}
where $f^{(n)}(z)$ is the $n$th iterate of $f_{z}$. (Note that $f(z)$, similarly to the unitary transformation $U$, is determined by $\varepsilon$ and $\ket{\psi_{\rm ref}}$.) It can be shown~\cite{KalmanPRA2018} that for any reference state $\ket{\psi_{\rm ref}}$ and $\varepsilon$, the nonlinear $f(z)$ transformation can be decomposed into the combination of M\"{o}bius transformations (corresponding to single-qubit rotations into the direction of the reference state $\ket{\psi_{\rm ref}}$) and a "core" nonlinear transformation $f_{\varepsilon}(z)=z^{2}/\varepsilon$, that only depends on $\varepsilon$. This essential, nonlinear part (without the further single-qubit rotations) corresponds to  matching the unknown state $\ket{\Phi_{0}}$ to the reference state $\ket{\psi_{\rm ref}}=\ket{0}$, and can be implemented by performing the protocol with the unitary
\begin{equation}
    U_\varepsilon = \begin{pmatrix}
      \varepsilon & -\frac{1}{\sqrt{2}}\sqrt{1-\varepsilon^2} & \frac{1}{\sqrt{2}}\sqrt{1-\varepsilon^2} & 0 \\
      0 & \frac{1}{\sqrt{2}} & \frac{1}{\sqrt{2}} & 0\\
      0 & 0 & 0 & 1\\
      \sqrt{1-\varepsilon^2} & \frac{1}{\sqrt{2}}\varepsilon & -\frac{1}{\sqrt{2}}\varepsilon & 0
    \end{pmatrix},
    \label{eq:Ueps}
\end{equation}
in every iteration. Figure~\ref{fig:circ_full} shows the corresponding quantum circuit implementation of the protocol for $n=2$ iterations. 

By using the parameterization $z=e^{i\phi_0}\tan(\theta_0/2)$ (where $\theta_{0}$ and $\phi_{0}$ are the spherical Bloch-sphere coordinates of $\ket{\Phi_{0}}$), the state of the single, final kept qubit after $n$ iterations can be written as
\begin{equation}
    \ket{\Phi_{n}}\!=\!\mathcal{N}_{n}\!\left[\varepsilon^{2^{n}\!-1}\!\left(\!\cos\frac{\theta_0}{2}\!\right)^{\!\!2^{n}}\!\!\ket{0} + e^{i2^{n}\phi_{0}}\!\left(\!\sin\frac{\theta_0}{2}\!\right)^{\!\!2^{n}}\!\!\ket{1}\!\right]\!\!.
    \label{eq:Phi_n}
\end{equation}

Let us note that when implementing the circuit of Fig.~\ref{fig:circ_full} we will intentionally use our own decomposition of the $U_\varepsilon$ operation, so that we can have a test protocol that uses the circuit with the same number of CNOT gates in all of the devices and that we have the possiblity of analyzing the results on the model level.   
The decomposition of $U_\varepsilon$ into CNOT gates and single-qubit unitaries can be achieved using our previous description~\cite{OrtegaPS2023}. It can be shown that $U_\varepsilon$ can be decomposed into two CNOTs and 6 single-qubit gates, following the procedure described in~\cite{TucciArxiv2005} and the results in~\cite{VidalPRA2004}. 
We note that if one intends to use our protocol to test quantum computers from different providers, one can leave the transpilation of the circuit to the local gate set of the given device to be performed by the the provider. 

\section{Methods}
\label{sec:statframe}

In the NISQ era, the impact of noise and errors in the devices are of special importance. 
In~\cite{ReschACM2021} a wide spectrum of error sources have been identiﬁed, ranging from environmental interactions, qubit interactions, imperfect operations
and detection errors, that can all contribute to deviations from the ideal result. On the other hand, since the outcome of quantum measurements are inherently probabilistic, there is always an inevitable statistical uncertainty in the estimation of a given quantity (e.g., the probability of an output bit string).  

Here we consider the following quantity: The success probability of performing $n$ iterations of the protocol described in Sec.~\ref{sec:implprot}. Its ideal value for a given input state (parameterized by the angles $\theta_{0}$ and $\phi_{0}$) can be written as
\begin{equation}
    p_{s}^{(n)} 
    = \varepsilon^{2^{n+1}-2}\left(\cos\frac{\theta_0}{2}\right)^{2^{n+1}} + \left(\sin\frac{\theta_0}{2}\right)^{2^{n+1}}.
    \label{eq:ps}
\end{equation}
It is important to note here that $p_{s}^{(n)}$ is independent of $\phi_{0}$, thus, in the case of an ideal (noise-free) quantum circuit, its estimated value, $p_{s}^{(n)}$ should not vary (apart from statistical uncertainty) as a function of $\phi_{0}$. We will use this fact as a reference to identify possible error sources in the tested quantum computers in Sec.~\ref{sec:heur}. 

In order to distinguish statistical errors from device errors, we will rely on the fact that the statistical uncertainty (which is characterized by the standard deviation $\sigma_s = \sqrt{p_s(1-p_s)/M}$) decreases as $1/\sqrt{M}$ with the number of measurements $M$. However, device noise is not expected to significantly change with $M$, therefore, by applying a large enough number of measurements (i.e., shots and runs)~\cite{ParisLNP2004}, so that statistical uncertainty becomes very small, one can, with high probability, tell apart deviations that were caused  mainly by the errors present in the device.  



\section{Results}
\label{sec:res}
\begin{figure}[htb]
  \centering
  \includegraphics[scale=0.35]{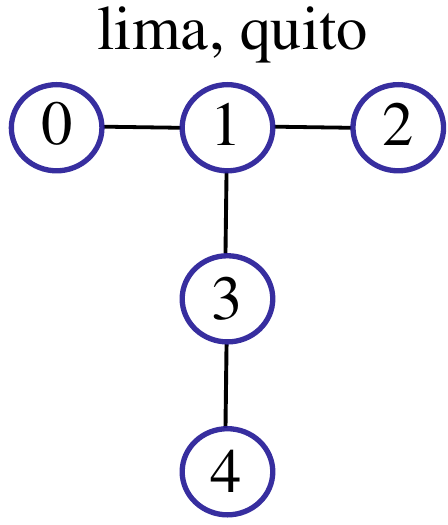}
   \includegraphics[scale=0.35]{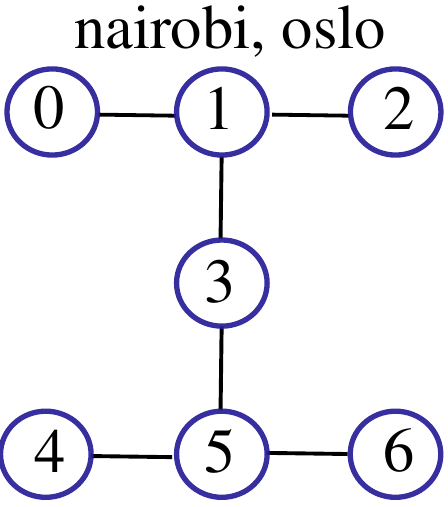}\\ \vspace{1ex}
   \includegraphics[scale=0.35]{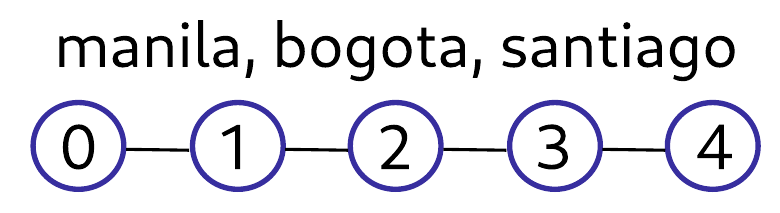}
  \caption{Topology of the tested devices.}
  \label{fig:topology}
\end{figure}

We implemented our test protocol corresponding to $\varepsilon\approx0.973$ on freely available IBM superconducting quantum computers~\cite{IBMqcs}. Their respective topologies are shown in Fig.~\ref{fig:topology}. We note that we used  neighboring qubits for the tests in the devices, in order to avoid unnecesary SWAP gates, and performed the post-selection at the very end of the circuit as mid-circuit mesurements were not available in the devices. At the time of writing, all these devices are already retired. The details about the execution dates of the circuits can be found in App.~\ref{sec:app_dates}.

\begin{figure}[htb]
  \includegraphics[width=0.855\columnwidth]{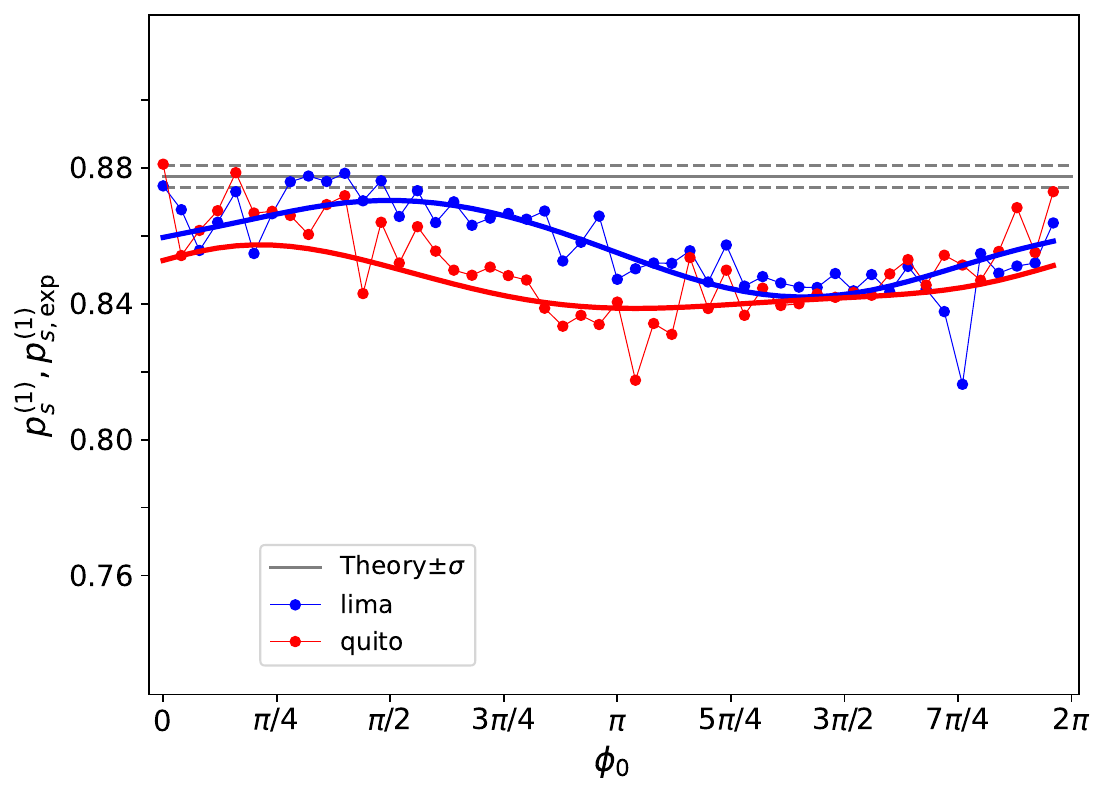}
  \includegraphics[width=0.855\columnwidth]{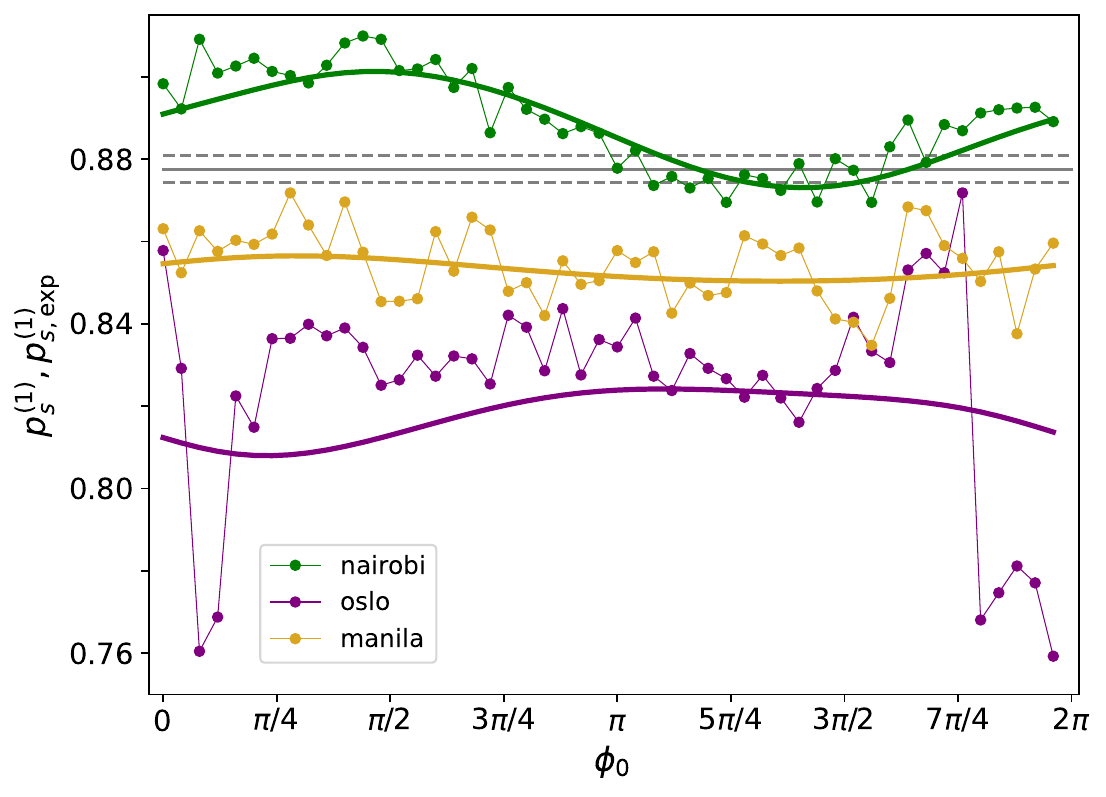}
  \caption{\label{fig:res1step} The success probability corresponding to one iteration of the test protocol (on qubits labelled 0 and 1 in all of the devices) as a function of the initial angle $\phi_{0}$. Colored dots represent the average estimated success probability for the different IBM quantum computers (shown in the insets), obtained by averaging 5 consecutive estimations of $p_{s}$ from experiments run with 2000 shots each. The solid and dashed grey lines correspond to the theoretical success probability (in this case $p_{s}^{(1)}=0.8775$) and its standard deviation, corresponding to $M=5\times2000$ experiments, respectively. The colored solid curves show a fit of the points with an error model discussed in Sec.~\ref{sec:heur}. Thin colored lines are guide to the eye.}
\end{figure} 

\begin{figure}[htb]
  \centering
  \includegraphics[width=0.855\columnwidth]{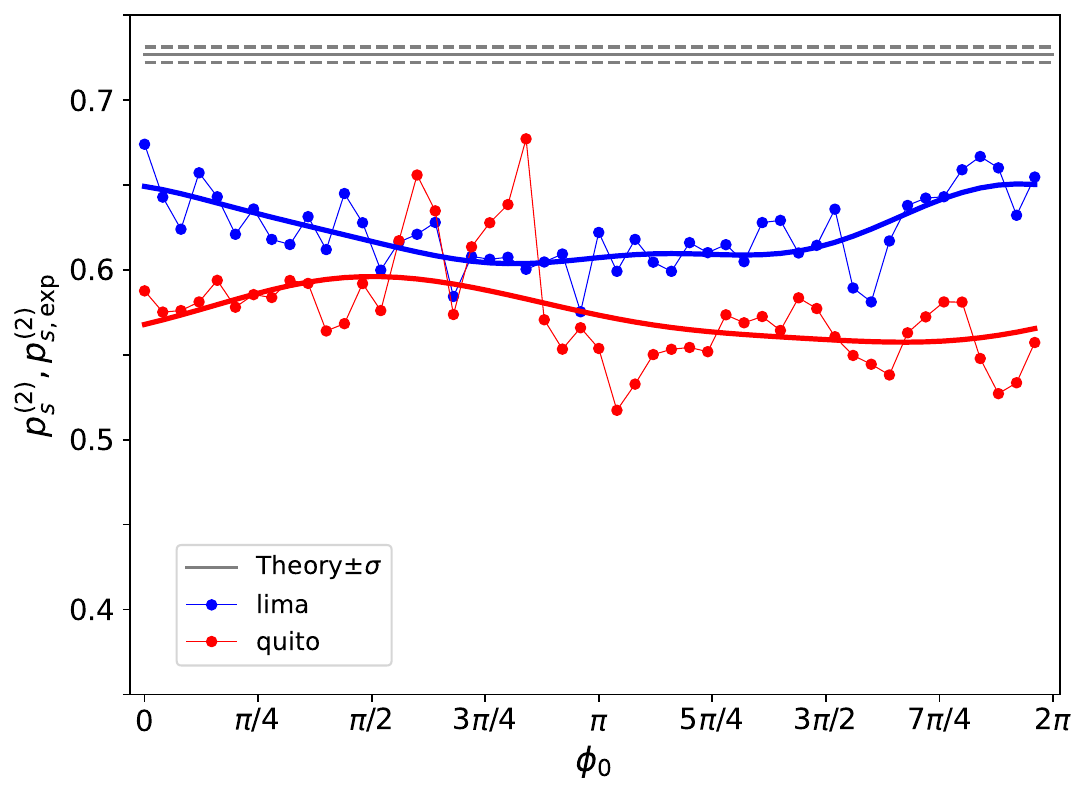}
  \hspace{0.33cm}
  \includegraphics[width=0.855\columnwidth]{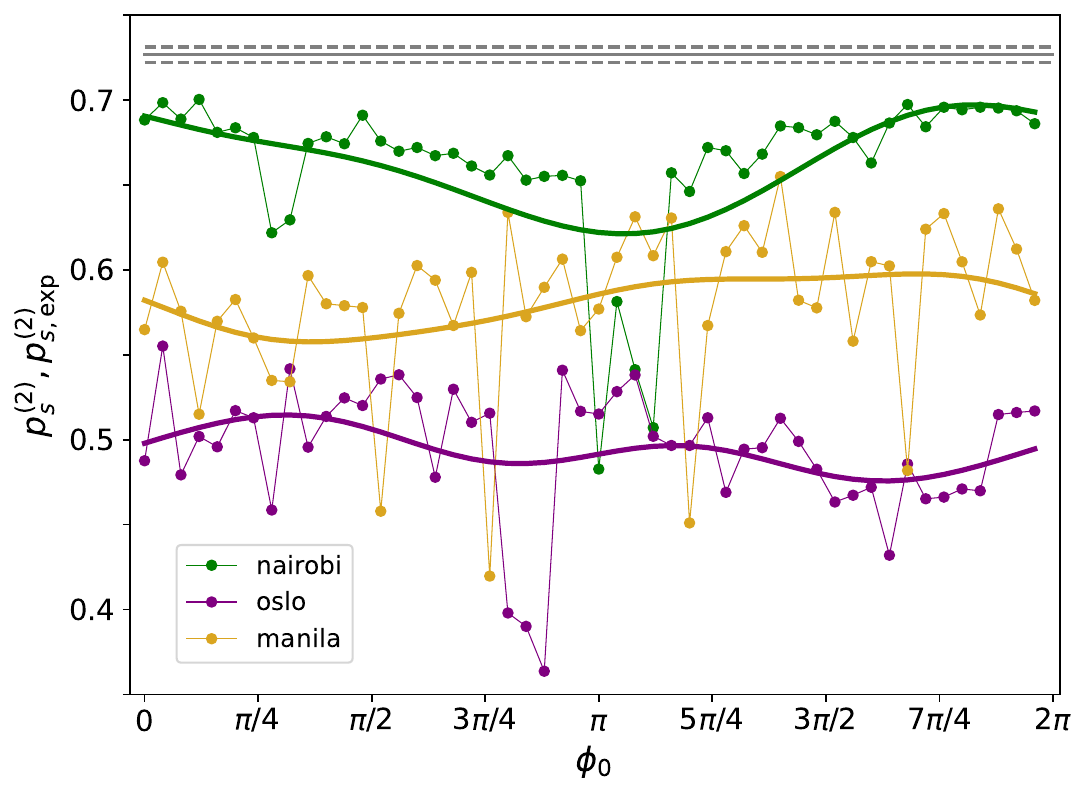}
  \includegraphics[width=0.855\columnwidth]{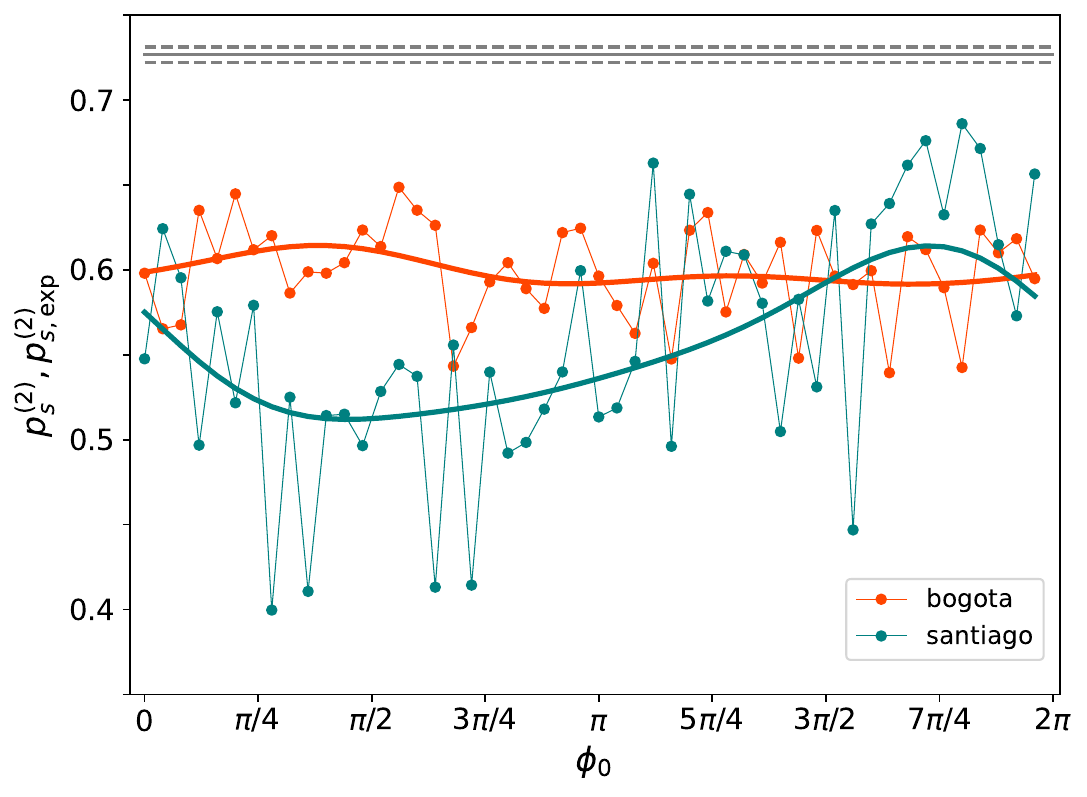}
  \caption{The success probability corresponding to two iterations of the test protocol as a function of the initial angle $\phi_{0}$. Solid and dashed grey lines show the theoretical success probability  $p^{(2)}_{s}=0.7267$ and the corresponding standard deviation, respectively. The experiments and the statistical analysis were carried out in the same way as in Fig.~\ref{fig:res1step}. The circuits were run on qubits $(0,1,3,4)$ in \latdev{lima} and \latdev{quito}; $(0,1,2,3)$ in \latdev{bogota}, \latdev{santiago} and \latdev{manila}; and on $(0,1,3,5)$ in \latdev{nairobi} and \latdev{oslo}. The colored solid curves show a fit of the points with an error model discussed in Sec.~\ref{sec:heur}.}
  \label{fig:res2step}
\end{figure}

Figures~\ref{fig:res1step} and \ref{fig:res2step} show the ideal and experimentally estimated success probabilities for one and two iterations of the protocol, respectively, for 50 different values of the initial angle $\phi_{0}$. For each value of $\phi_{0}$, the circuits were run 5 times consecutively (in batches, without significant time delay or any calibration in between), each with 2000 shots. Therefore, we can assume that the environmental conditions were essentially the same during all 10000 measurements, thus we may consider them together for our statistical analysis. 
The solid gray line correspond to the theoretical $p_s^{(n)}$, while the dashed gray lines represent the theoretical statistical standard deviation, corresponding to the number of experiments taken (i.e., $M=10000$). For each device, the initial state was prepared with an angle $\theta_0=\pi/8$. 

It can be seen that the experimentally estimated success probabilities (represented by the dots) lie significantly outside the statistical tolerance for most values of $\phi_{0}$, indicating the presence of device errors. Moreover, the results tend to follow a systematic  trend as a function of $\phi_{0}$ (see the results for \latdev{lima}, \latdev{quito}, \latdev{nairobi}, and \latdev{oslo}), which is in contrast with what we would expect from theory (grey line). The experimentally estimated success probability is lower than the theoretical value except for the case of \latdev{nairobi}, where many of the data points lie higher than $p_{s,\text{id}}$. Already from the figures one can see that \latdev{lima}, which has a QV of 8 performes better than \latdev{quito}, which has a QV of 16.  By comparing the upper and lower figures one can see that \latdev{lima} is also comparable in performance with \latdev{manila}, which has a QV of 32.

\begin{figure}[htb]
  \centering
  \includegraphics[width=0.855\columnwidth]{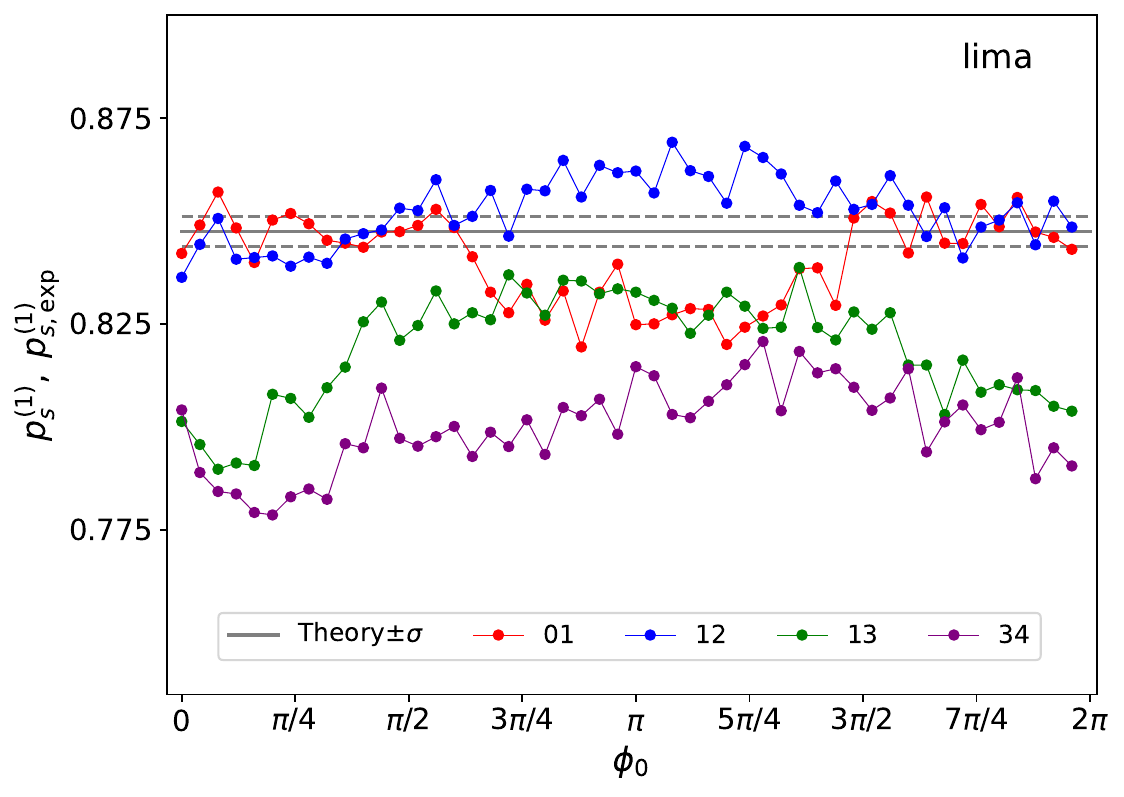}
  \includegraphics[width=0.855\columnwidth]{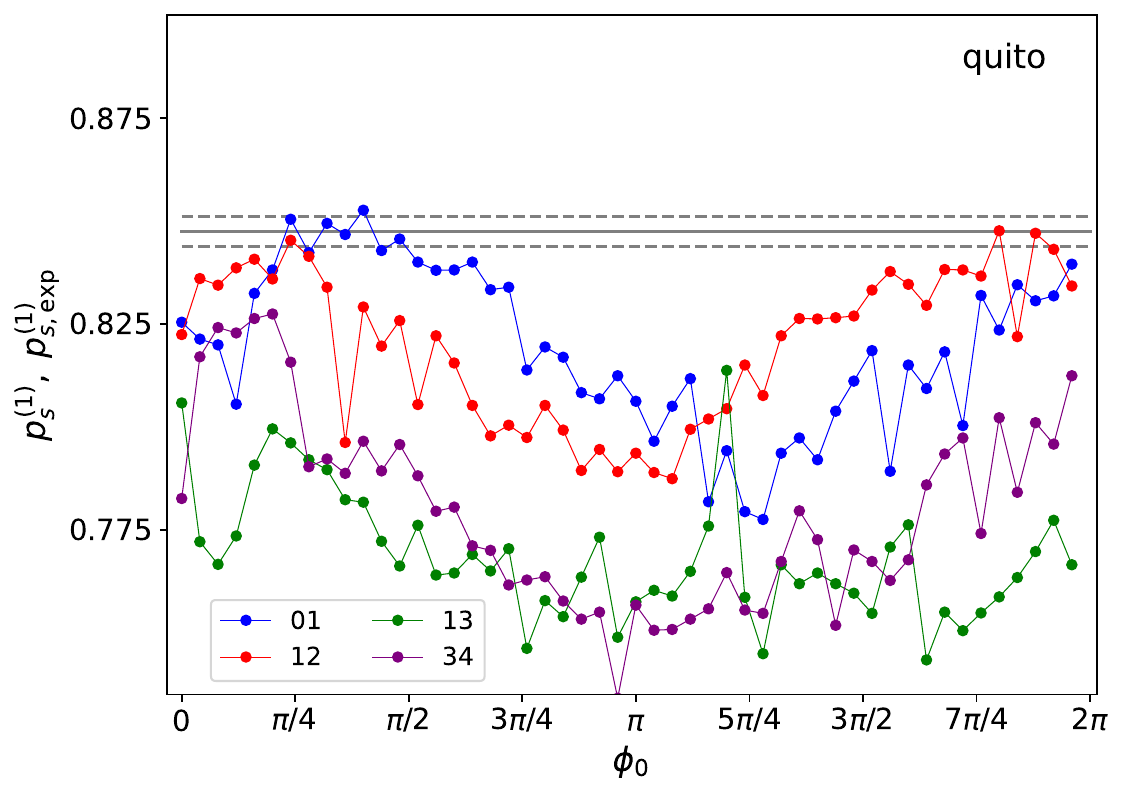}
  \includegraphics[width=0.855\columnwidth]{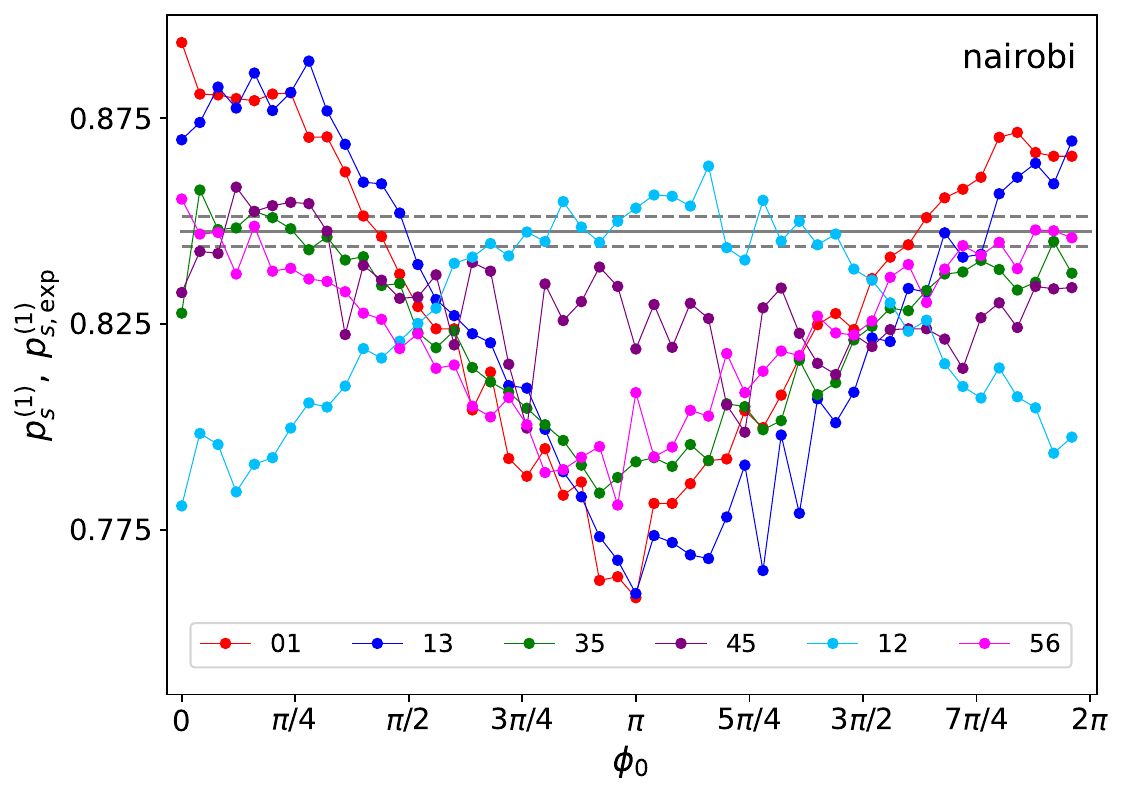}
  \caption{The success probability corresponding to one iteration of the test protocol as a function of the initial angle $\phi_{0}$ for \latdev{lima}, \latdev{quito} and \latdev{nairobi}. The input angle was set to $\theta_0\approx 0.8775$, for which the theoretical success probability is $p^{(1)}_s \approx 0.8474$ (solid grey line). The  dashed grey lines show the corresponding standard deviation. The experiments and the statistical analysis were carried out in the same way as in Fig.~\ref{fig:res1step}, but the combinations of qubits were varied (see the legend on the figures).}
  \label{fig:res1ps_outsideeps}
\end{figure}

\begin{figure}[htb]
  \centering
  \includegraphics[width=0.855\columnwidth]{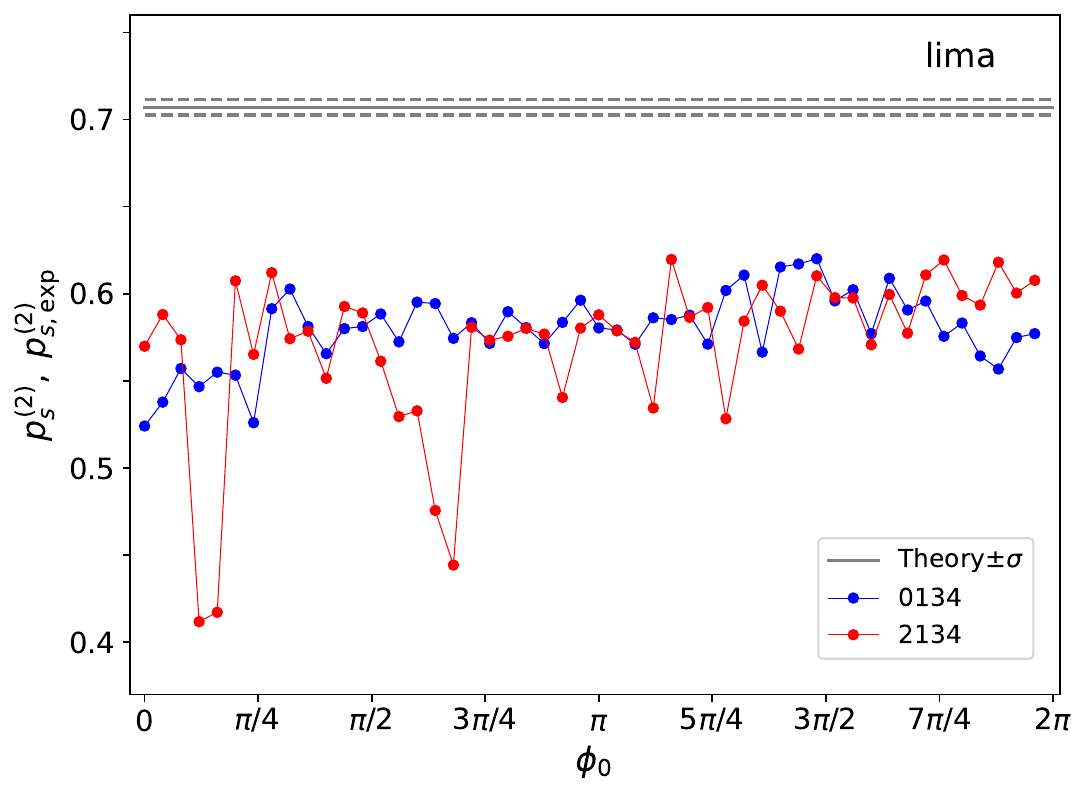}
  \includegraphics[width=0.855\columnwidth]{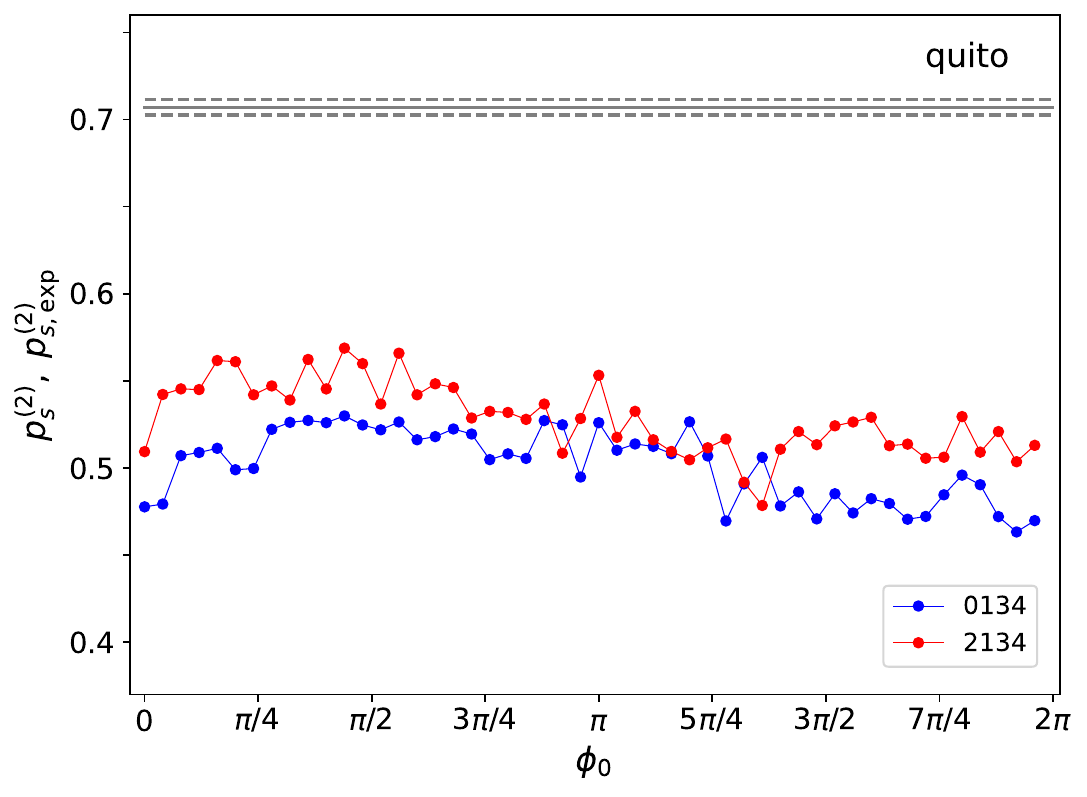}
  \includegraphics[width=0.855\columnwidth]{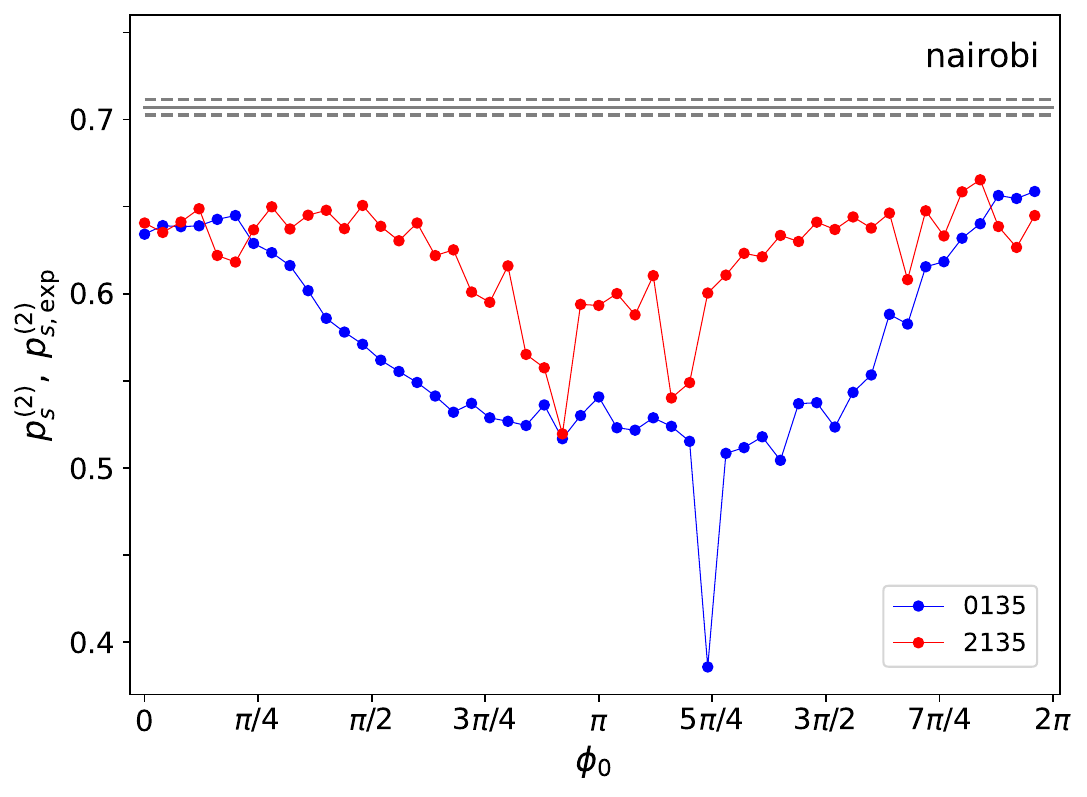}
  \caption{The success probability corresponding to two iterations of the test protocol as a function of the initial angle $\phi_{0}$ for \latdev{lima}, \latdev{quito} and \latdev{nairobi}. Solid and dashed grey lines show the theoretical success probability and the corresponding standard deviation, respectively. The experiments and the statistical analysis were carried out in the same way as in Fig.~\ref{fig:res1step}, but the combinations of qubits were varied (see the legend on the figures).}
  \label{fig:res1ps_outsideeps_twosteps}
\end{figure}


We have also tested other combinations of qubits on three quantum computers. Figures~\ref{fig:res1ps_outsideeps} and \ref{fig:res1ps_outsideeps_twosteps} show these results for one and two iterations of a slightly different scheme on \latdev{lima}, \latdev{quito}, and \latdev{nairobi}. Note that they all have different quantum volumes (\latdev{lima} QV8, \latdev{quito} QV16 and \latdev{nairobi} QV32). In these cases we used a different initial angle $\theta_0\approx 2.5571$, for which the theoretical success probability is $p^{(1)}_s \approx 0.8474$. It can be seen that in \latdev{lima} and \latdev{quito}, the qubit combinations (01) and (12) yielded results closer to the theoretical success probability, while the other combinations generally performed worse on our test. By looking at the calibration data   published by IBM for these qubit combinations on the dates that the experiments were run, we could see that the error of the CNOT gate is smaller between these qubit pairs, than in the other cases, suggesting that  the CNOT-error can have a significant contribution to the overall deviation from the theoretical value. The systematic variation of the data points, which we have seen on Fig.~\ref{fig:res1step} is visible here as well. \latdev{nairobi}, which has more qubits (thus more combinations) shows a rather similar behavior for most combinations, except for (45) and (12). The former combination does not produce significant systematic variations, while the latter one follows a trend opposite to the other ones. 

The test corresponding to two iterations of the protocol was implemented on two different combinations of the qubits on all three devices. The results are shown in Fig.~\ref{fig:res1ps_outsideeps_twosteps}. Note that the 4 chosen qubits are always neighboring ones. The previously seen systematic variation of the data points is not so much visible any more, except for the case of \latdev{nairobi}, where one can also see that replacing qubit 0 with qubit 2 leads to a significant increase of the success probability in the $\phi_0\in\left[\pi/4,7\pi/4\right]$ range. This effect can also be attributed to the smaller CNOT-error between qubits (21) compared to that between qubits (01). We note that the difference between the CNOT-errors of qubits (01) and (21) are smaller in the case of \latdev{lima} and \latdev{quito}, as one would expect. Overall, one can observe, that the experimentally estimated success probabilities are much lower than the theoretical one, compared to the case when the one-iteration test was applied (see Figs.~\ref{fig:res1ps_outsideeps}). 


\section{Benchmark metrics}
\label{sec:benchmark}

When defining a benchmark metric based on our test, we intend to follow the usual approach of the community and give an easily comprehensible score, or scores, that can help a high-level user to compare the performance of different quantum computers. 
There are several reasons to take this approach: (1) currently the majority of users do not have a low-level access (such as device tuning, manipulation of pulses, optimize transpilation times,  etc.) to current quantum devices, so they must rely only on the day-to-day (automatic) calibration; (2) queues are long and waiting times can affect the results when an analysis in a range of parameters is required; (3) in the long run, users want to implement their circuits without worrying on low-level details of the implementation. 


We have argued above (see Sec.~\ref{sec:statframe}) that if the number of experiments (i.e., shots) is sufficiently large so that the corresponding statistical uncertainty (i.e., standard deviation) is sufficiently small, then device errors can be distinguished from pure statistical fluctuations (see e.g. Refs.~\cite{GilyenArxiv2021, ZimborasQuantum2022}). We could see from Figs.~\ref{fig:res1step}-\ref{fig:res1ps_outsideeps_twosteps} that a large percentage of experimental data indeed falls outside the $p_{s}^{(n)}\pm\sigma_{s}^{(n)}$ interval, suggesting the presence of device errors. (Note that the probability of findig a result in this interval is $68\%$, where $\sigma_{s}^{(n)}$ is the standard deviation of the binomial (Gaussian) distribution with mean $p_{s}^{(n)}$.)

\begin{figure}[htb]
  \centering
  \includegraphics[width=0.98\columnwidth]{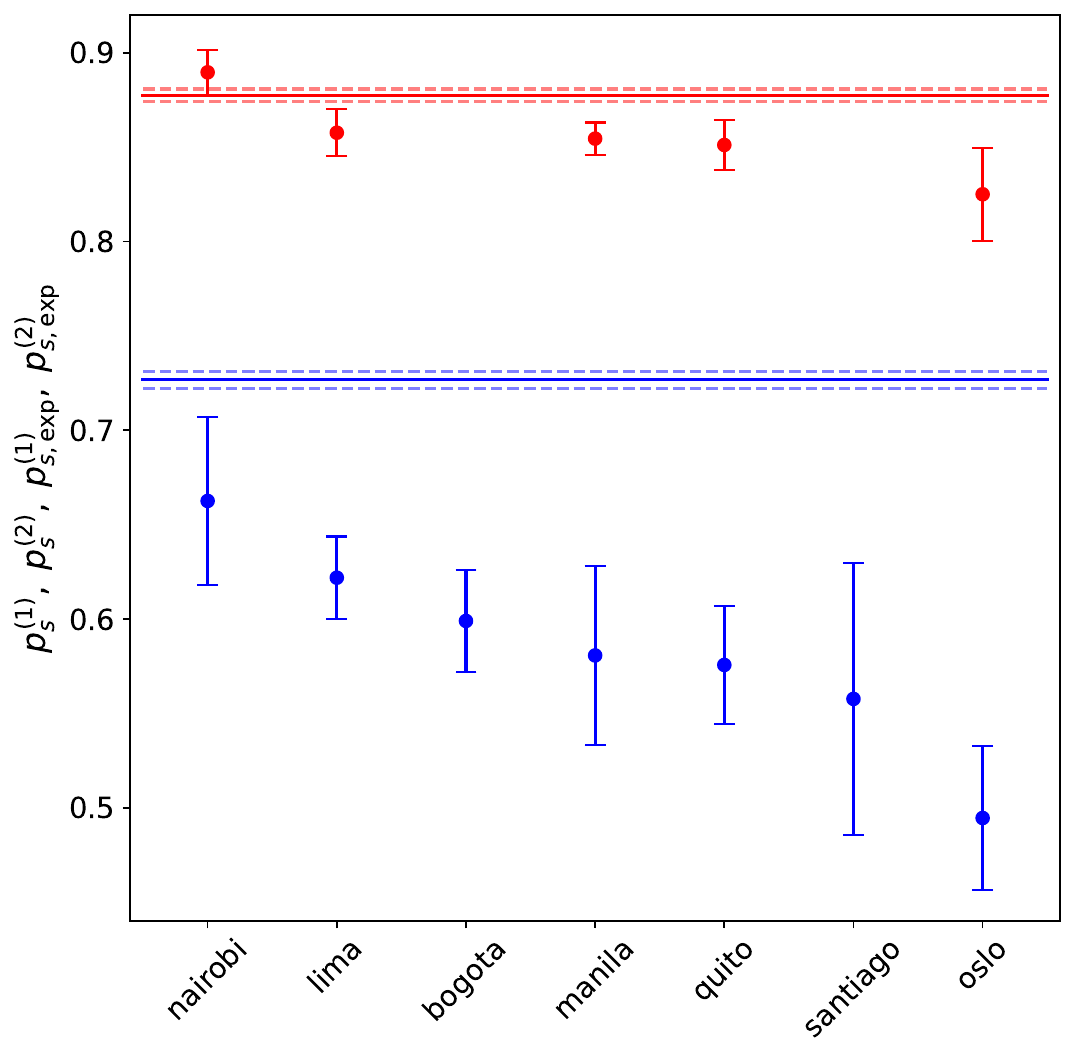}
  \caption{The average of the experimentally estimated success probabilites and the corresponding standard deviations for each of the tested devices for tests with $n=1,2$ iterations of the protocol. The theoretical success probabilities and standard deviations are also shown for reference. The missing data points mean that the test was not performed in the given device.}
  \label{fig:meanvar}
\end{figure}

Let us recall that Figs.~\ref{fig:res1step} and \ref{fig:res2step} show the experimentally estimated success probabilities which were obtained by running the test circuits with 2000 shots repeated 5 times (altogether $10^4$ shots) for each value of $\phi_0$, for $50$ different values of $\phi_0$. Since the theoretical success probabilities $p_{s}^{(n)}$ are independent of $\phi_{0}$, but the $p_{s,\text{exp}}^{(n)}(\phi_{0})$ are not, it is straightforward to compare the average of the latter over $\phi_{0}$ -- which we denote by $\bar{p}_{s,\rm exp}^{(n)}$ -- to the ideal, theoretical $p_{s}^{(n)}$, and the experimentally observed  standard deviation $\sigma_{s,\text{exp}}^{(n)}$ to its ideal counterpart $\sigma_{s}^{(n)}$. We define therefore the following two metrics:
\begin{align}
F^{(n)}&=1-\frac{|\bar{p}_{s,\rm exp}^{(n)}-p_s^{(n)}|}{p_s^{(n)}} \,, \\
S^{(n)}&=\frac{\sigma_{s,\rm exp}^{(n)}}{\sigma_s^{(n)}}\,,
\end{align}
where
$\sigma_{s,\text{exp}}^{(n)}=\sqrt{\frac{1}{50}\sum_{i=1}^{50}\left(p_{s,\rm exp}^{(n)}(\phi_{0,i})-\bar{p}_{s,\rm exp}^{(n)}\right)^{2}}$.
$F^{(n)}$ is a quantity which represents how well the success probability could be estimated with our test using $n$ iterations of the protocol. Note that $0\leq F^{(n)}\leq 1$, where $F^{(n)}=1$ would mean that $p_{s}$ could be estimated correctly by running the test. $S^{(n)}$, on the other hand, measures how much more the experimental estimates were spread around their average value than one would expect from pure statistical uncertainty, and as such, it intends to capture the overall noise that affects a given quantum computer.  
Tables~\ref{tab:scores1} and  \ref{tab:scores2} show these two metrics calculated for the tested devices based on the data of Figs.~\ref{fig:res1step} and \ref{fig:res2step}, respectively, together with the corresponding quantum volumes of the devices, while Fig.~\ref{fig:meanvar} shows $\bar{p}_{s,\rm exp}^{(n)}$ and the theoretical success probability, as well as the corresponding standard deviations for each of the tested devices for tests with $n=1,2$ iterations of the protocol.

\begin{table}[h]
    \begin{tabular} {|c|c|c|c|c|}
        \hline
        Device & QV & $\bar{p}_{s,\rm exp}^{(1)}/p_s^{(1)}$ & $F^{(1)}$ & $S^{(1)}$ \\ \hline
        nairobi & 32 &1.014 & 0.986 & 3.609\\ 
        lima & 8 & 0.977 & 0.977 & 3.812 \\ 
        manila & 32 &0.974 & 0.974 & 2.599 \\
        quito & 16 &0.970 & 0.970 & 4.041 \\ 
        oslo & 32 &0.940 & 0.940 & 7.544 \\
        \hline
    \end{tabular} \\
    \caption{Benchmark metrics calculated from the test results corresponding to one iteration of the protocol (see  Fig.~\ref{fig:res1step}). The quantum volume of the tested devices is also displayed for reference.}
    \label{tab:scores1}   
\end{table}
\begin{table}[h]
    \begin{tabular} {|c|c|c|c|c|}
        \hline
        Device & QV & $\bar{p}_{s,\rm exp}^{(2)}/p_s^{(2)}$ &  $F^{(2)}$ &  $S^{(2)}$\\ \hline
        nairobi & 32 & 0.912 & 0.912 & 9.953 \\ 
        lima & 8 & 0.856 & 0.856 & 4.893\\ 
        bogota & 32 & 0.824 & 0.824 & 6.077\\
        manila & 32 & 0.799 & 0.799 &  10.650\\
        quito & 16 & 0.792 & 0.792 & 7.045\\ 
        santiago & 32 & 0.767 & 0.767 &  16.204\\
        oslo & 32 & 0.681 & 0.681 & 8.539\\
        \hline
    \end{tabular} \\
    \caption{Benchmark metrics calculated from the test results corresponding to two iterations of the protocol (see  Fig.~\ref{fig:res2step}). The quantum volume of the tested devices is also displayed for reference.}
    \label{tab:scores2}   
\end{table}

It can be seen that the device which produces the highest $F^{(n)}$ for both tests is \latdev{nairobi}, although the standard deviation of the errors is not the lowest among the other tested devices. \latdev{lima}, which is ranked 2nd on our test (both for $n=1$ as well as $n=2$ iterations) performs quite consistently also if we  consider the metrics $S^{(n)}$, even though its QV is the lowest among the devices. One can also observe that the ranking of the devices based on our metrics $F^{(n)}$ (or $S^{(n)}$) does not correlate with the QV of the devices. One possible reason for this can be that our test is based on a specific quantum circuit, and not many randomly chosen ones, like in the case of the QV benchmark. Another possible explanation can be that in Tables~\ref{tab:scores1} and  \ref{tab:scores2} we presented test results for a single combination of the qubits, and as such, we do not average all possible errors. Still, we believe that our test, toghether with the above defined metrics can be an informative tool for a high-level user to decide which sets of qubits possess the desired quality to run a given quantum circuit.

\section{Estimation of error parameters}
\label{sec:heur}

Our goal in this section is to estimate errors of the quantum computers, revealed by the careful analysis of the results of executing our nonlinear protocol. The results that we obtained by fixing $\theta_0$ and varying $\phi_0$ (Figs.~\ref{fig:res1step}, \ref{fig:res2step}, \ref{fig:res1ps_outsideeps}) deviate from the expected constant value $p_s^{(n)}$. The mechanism behind the shifting of the average value can be related to two typical types of errors: one modelled by a depolarizing channel and the other described by amplitude damping. In addition to the shifted average value, we can observe in many cases a slowly changing functional variation with respect to $\phi_0$ (Figs.~\ref{fig:res1step} \ref{fig:res2step}, \ref{fig:res1ps_outsideeps}). We intend to explain the $\phi_0$-dependent variations by considering coherent errors, known to occur in this type of quantum devices and quantum computers in general~\cite{MakhlinRMP2001}, \cite{ClarkeNat2008}, \cite{KrantzAPR2019} and~\cite{ReschACM2021}. Specifically, we propose and analyze a digitalized error model~\cite{ShorPRA1995}, \cite{AruteNat2019}. Such error models are routinely applied in current NISQ systems, see for instance~\cite{TroutNJP2018}, \cite{ZhuPRA2021}. Here we aim at identifying only some dominant sources of errors providing a plausible explanation of the observed test results. We will show that this type of test allows for a quantitative estimation of error parameters, or a combination thereof. 

The effect of depolarizing channels, which may be combined to one effective depolarizing channel, lowers the average measured probability seen in Figs.~\ref{fig:res1step}, \ref{fig:res2step}, see Appendix~\ref{sec:depchannel}. The amplitude damping channel, on the other hand, increases the average measured probability, see Appendix~\ref{sec:ampdamp}. The overall shift of the average measured probability will therefore depend on the combination of these channels.


In what follows, we argue that in order to describe the results in Figs.~\ref{fig:res1step} and \ref{fig:res2step} one needs to consider essentially the following processes: (1) the combined contribution of a depolarizing channel and an amplitude damping channel (2) coherent gate errors. 
We mention in passing that there might be other types of errors which can shift the measured  success probability. For instance, the error of measuring state $|1\rangle$ while $|0\rangle$ was prepared and vice versa can both increase or decrease the measured success probability, depending on the concrete value of the readout error probabilities as well as the success probability. A quick derivation shows that, for the parameters described in the specific device cards of the devices tested, the shift caused by the measurement errors is significantly smaller than the effect of the amplitude damping channel or the depolarizing channel.

\subsection{Coherent error model}
\label{sec:cohmod}

The characterization of coherent gate errors in quantum computers is a nontrivial task, although, their knowledge is of high importance, since coherent errors, in principle, can be corrected by appropriately modifying the control pulses in the devices. 

It is clear that the type of results that we obtain by fixing $\theta_0$ and varying $\phi_0$ contain a tendency that deviates from a constant $p_s^{(n)}$, see Figs.~\ref{fig:res1step} and \ref{fig:res2step}. Furthermore, they also display a clear systematic dependency on $\phi_{0}$ that needs to be understood. In this section, we propose a simple model to reproduce the ``oscillations'' observed in the experimental results by assuming the presence of coherent single- and two-qubit gate errors.

After the transpilation procedure, for all the quantum computers used here, the resulting circuit is a combination of the native gates $\sqrt{{\rm X}}$, $R_z$ and CNOT. We aim at constructing a model that takes into account misrotations (or usually called ``over/under-rotations'') in these quantum gates only. As it is common in the literature, one can consider that there is only one type of single-qubit error that is present in the $\sqrt{X}$-gate, while the $R_z$ rotation is considered ``virtual''~\cite{mckayPRA2017} and as such, error free. In this way, the device card of a certain quantum computer from IBM only reports an error different from zero for $\sqrt{X}$ gates over different qubits. The gate $\sqrt{X}$ is a physical rotation over the $x$-axis of the Bloch sphere by an angle of $\pi/2$, that is
\begin{equation}
    R_x(\pi/2) = e^{-i\pi/4}\sqrt{X},
\end{equation}
where, in what follows, the global phase of $e^{-i\pi/4}$ can be ignored. This means that a misrotation of the $\sqrt{X}$ gate can be taken into account as
\begin{equation}
    R_x(\pi/2 + \alpha) = R_x(\pi/2)R_x(\alpha),
\end{equation}
where $\alpha$ is a real number, i.e. the (small) error in implementing a $\sqrt{X}$-gate. In what follows, we will use a subindex $i$ to indicate that $\alpha_i$ is a misrotation in the $\sqrt{X}$ that is applied to qubit $i$.

In order to consider the two-qubit errors introduced by a CNOT gate, one needs to go deeper into the implementation of such a gate and look for the ``physical'' gate from which a CNOT is constructed. In all the devices used here, the CNOT gate is implemented by a cross-resonance gate (CR)~\cite{RigettiPRB2010}, ~\cite{AlexanderQST2020}, where
\begin{equation}
    U_{\rm CR}(\beta) = \exp(-i\beta Z_1 X_2)
\end{equation}
with $Z_1X_2 = Z_1 \otimes X_2$ and $\beta$ is real. If one fixes $\beta = \pi/4$, one finds that
\begin{equation}
    U_{\rm CR}(\pi/4) = e^{-i\pi/4}(Z_1 X_2)^{1/2}
\end{equation}
and the CNOT gate can be obtained as the matrix multiplication
\begin{equation}
    {\rm CNOT} \equiv (Z_1 I_2)^{-1/2}(Z_1 X_2)^{1/2}(I_1X_2)^{-1/2},
    \label{eq:CNOTquiv}
\end{equation}
where again a global phase of $e^{-i \pi/4}$ can be ignored. Similarly to the one-qubit case, the coherent errors can be introduced by writing $\beta$ as
\begin{equation}
    \beta = \pi/4 + \Lambda,
\end{equation}
where $\Lambda$ is a real number, that is, the misrotation angle. We will use $\Lambda_i$ to label the coherent error present when  ``coupling'' a given pair of qubits identified by $i$. The device card only shows the error in the CNOT gate, and there is no further information about the local unitaries $Z_1I_2$ and $I_1X_2$. Because of this, we shall consider that the error in a CNOT reported is directly linked with the implementation of the CR-gate. In any case, one qubit errors can be included in the corresponding $\sqrt{X}$-gate errors.

\subsection{The fitting procedure}
\label{sec:decmod}
\begin{figure*}
  \includegraphics[scale=0.8]{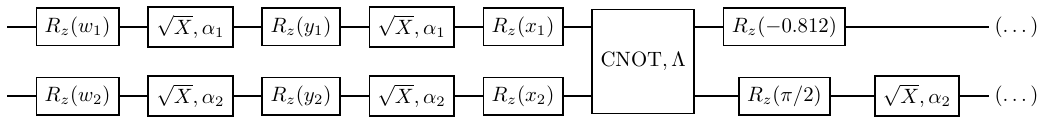}
  \includegraphics[scale=0.8]{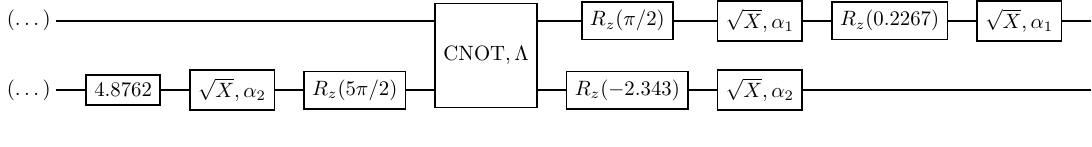}
  \caption{Transpiled circuit for one step of the protocol. For details see Sec.~\ref{sec:decmod} and App.~\ref{app:angles}}
  \label{fig:circtranspiled}
\end{figure*}
In the quantum computers that we tested, our full circuit (including state preparation)  is implemented as a combination of the native gates $\sqrt{{\rm X}}$, $R_z$ and CNOT gates, the latter being implemented by CR-gates. Figure~\ref{fig:circtranspiled} shows the generic circuit after the transpilation procedure of $U_\epsilon$ and state preparation for one step of the protocol. The values of the angles $x,y,z$ can be computed using the formulas of App.~\ref{app:angles}. Note that in this case $x = x(\epsilon, \theta_0,\phi_0)$ and similarly for $y$ and $z$. If $U$ denotes the circuit after $n$ steps (including state preparation), our whole procedure can be understood as
\begin{equation}
    U|0\rangle^{\otimes N} = (U_{11}, U_{21}, \dots, U_{N1})^T,
    \label{eq:Ugen1}
\end{equation}
where the initial experimental preparation of the $N$ qubits is the tensor product $|0\rangle^{\otimes N}$, $T$ denotes the transpose, and the application of $U$ to $|0\rangle^{\otimes N}$ has the result of obtaining a vector equal to the first column of $U$. As we are interested in the success probability $p_s^{(n)}$ after $n$-steps (see Eq.~\ref{eq:ps}), we only need two entries of Eq.~\ref{eq:Ugen1}, namely
\begin{equation}
    p_s^{(n)} = |U_{11}|^2 + |U_{j1}|^2 
\end{equation}
where $U_{j1}$ corresponds to the coefficient of the vector $|10\dots 0\rangle$. In actual experiments, $p_s^{(n)}$ is further modified by all the other types of errors present. In our description, the other main sources of errors are either a depolarizing or an amplitude damping channel, as well as coherent errors of the actual gates implemented. The equation for $p_s^{(n)}$ would describe the experimental data if we can find an estimate for $p_{\rm dep}$ and $p_{\rm ad}$ (probability of depolarizing and amplitude damping), and one- and two-qubit misrotations $\alpha_{i}$ and $\Lambda_{i}$, respectively. The fitting procedure is thus a four-step process:
\begin{enumerate}
    \item Find the parameter $p_{\rm dep}$ or $p_{\rm ad}$ (depending on the case) for the experimentally estimated success probability $p^{(n)}_{s,\rm exp}$ in a given device (see Figs.~\ref{fig:res1step} and \ref{fig:res2step}). This parameter in turn describes an overall shift in the measured success probability with respect to the theoretical value of Eq.~\ref{eq:ps}.
    \item Shift $p^{(n)}_{s,\rm exp}$ using the previous step. The shifted data will then  ``oscillate'' around the theoretical value of $p_s^{(n)}$.
    \item Find the value for the coherent errors $\alpha_{i}$ and $\Lambda_{i}$, using as a bound the interval $[-\alpha^{\rm tol}_{i}, \alpha^{\rm tol}_{i}]$ and $[-\Lambda^{\rm tol}_{i}, \Lambda^{\rm tol}_{i}]$. In this step, $U$ in Eq.~\ref{eq:Ugen1} is written using the transpiled circuit of Fig.~\ref{fig:circtranspiled} for one step and each relevant gate takes into account a corresponding misrotation, as described above. In the case of two steps, the transpilation is more involved but it follows the main layout presented in Fig.~\ref{fig:circtranspiled}.
    \item Shift back the experimental data and the obtained fit as in step 2.
\end{enumerate}

The solid lines in Figs.~\ref{fig:res1step} and \ref{fig:res2step} are obtained following the above procedure. In both cases we have used scipy's function \latdev{curve\_fit}. Table~\ref{tab:params_1step} and \ref{tab:params_2step} shows the values of the parameters fitted for the test performed with one or two iterations of the protocol. Note that in all of the cases we assumed that the error $\alpha_{i}$ in the gate $\sqrt{X}$  acting on qubit $i$ is the same throughout the whole circuit. Similarly, for a given pair of qubits for which a CNOT is applied, the error $\Lambda_{i}$ does not change if the pair of qubits is the same, see Fig.~\ref{fig:circtranspiled}. One can see that, under this procedure, in the single-iteration case we need exactly three angles that describe the misrotations: two that describe the single-qubit errors $\alpha_{i}$, and one that describes the error $\Lambda_{i}$ of the CNOTs applied to the two qubits.

For the two-iteration case, the situation is very similar, although some extra features need to be taken into account. First, regardless of the device used, the set of qubits is the same (only four qubits and always the same physical qubits). Second, in order for the second iteration to be applied, a SWAP operation took place in order to exchange the state onto neighboring qubits. The overall count for required parameters is then 8: 4 single-qubit misrotations $\alpha_{i}$, and 4 two-qubit misrotations $\Lambda_{i}$. This was confirmed using the transpilation for all the devices, because the SWAP operation introduces at most 2 extra coherent errors (recall that the control and the target qubits alternate when the SWAP gate is written as three CNOTs). The results for the fitted parameters along with the maximum angle error from the experiments are shown in Table~\ref{tab:params_2step}. 

It can be seen from Figs.~\ref{fig:res1step} and \ref{fig:res2step} that the fitting obtained by this rather simple coherent error model gives an acceptable description of the deviations from the theoretical constant value. We note however, that for the quantum computer \latdev{oslo}, the fitting may not be fully reliable, because of a few, extremely outlying data points. In all the other cases the fitted function follows the tendency of the data points fairly well. Let us note that by the time our coherent error model was fully established, the devices have all been retired, hindering the possibility of further experimenting using pulse-level control to confirm our fitted values.

\section{Summary and outlook}

We investigated the performance of different freely available superconducting quantum computers offered by IBM. We used the so-called quantum state matching protocol~\cite{KalmanPRA2018,OrtegaPS2023} to test the devices, performing one and two iterations of the scheme. Our figure of merit was the overall success probability, which is theoretically invariant under the variation of the initial angle $\phi_0$ if $\epsilon$ and $\theta_0$ are fixed. We found that the experimentally estimated success probability shows a dependence on $\phi_0$  in almost all of the tested quantum computers. We analyzed the results from two different points of view. In the first case, we defined a benchmark metric that allowed us to score the exectution of the same circuits across different quantum devices, finding that the best device was \latdev{nairobi} which has a QV of 32. Interestingly, the scores obtained using our approach do not necessarily correlate with a given QV. We found cases in which, although the quantum volume was quite small, our benchmark score was relatively high. For instance, the device \latdev{lima} with QV 8 was the second best quantum computer according to our ranking. On the other hand, \latdev{oslo}, which has a QV of 32, and was released roughly at the same time as \latdev{nairobi},  still performed the worst of all the tested devices.

The second point from which we analyzed the results was the possible presence of coherent errors in the devices. In order to model this, we included coherent misrotations in the native gates and applied a fitting procedure. We obtained a fairly good fit to the experimental data using only a few set of parameters. 

Regarding the characteristics of our benchmark, our metrics can be compared to statistical performance and there is a clear advantage in using our state matching protocol as a test as it can reveal nontrivial coherent errors. Further advantages of our test protocol are: (1) we can test an exponentially large number of qubits by increasing the number of steps in the protocol, that is, $n$ iterations test $2^n$ qubits and (2) we can evaluate classically and efficiently the final theoretical expected result after the application of $n$ iterations of the protocol, as it only amounts effectively to the evaluation of the complex power function $f(z)=z^2/\epsilon$. In our analysis, we did not fully take advantage of another important feature of our test protocol, namely, its adaptivity. Here we only tested quantum computers up to 7 qubits, which only made possible the implementation of two iterations. On larger quantum computers, more iterations could be performed, and in order to maintain a high enough number of counts in the final, post-selected bit strings, we could choose a different $\varepsilon$ and input angle $\theta_{0}$. The structure of our protocol also makes it possible to test mid-circuit measurements and feed-forward in quantum computers. This type of protocol by construction provides a certain flexibility to check the quality of individual qubits or sets of qubits, and by carefully changing a few of the qubits on which we run the test circuits, we might also be able to reveal cross-talks inside the device.

\section*{Acknowledgements}
We thank Mikel Sanz and Arturo Acuaviva for pointing out reference~\cite{SmithCommACM1988}. This research was supported
by the Ministry of Culture and Innovation, and  the National Research, Development and Innovation Office within the Quantum Information National Laboratory of Hungary (Grant No. 2022-2.1.1-NL-2022-00004). OK and TK acknowledge support from the National Research, Development and Innovation Office of Hungary, project No. TKP-2021-NVA-04. OK acknowledges support from OpenSuperQPlus100. A. O. acknowledges support from the program ``Apoyos para la Incorporaci\'on de
Investigadoras e Investigadores Vinculada a la Consolidaci\'on Institucional de Grupos de Investigaci\'on 2023'' from CONAHCYT, M\'exico.

\appendix

\section{Depolarizing channel}
\label{sec:depchannel}
One of the main sources of errors in current quantum computers can be modelled with the depolarizing channel
\begin{equation}
    \mathcal{E}_{\rm dep}(\rho) = \frac{p_{\rm dep}}{D}\mathbb{1} + (1-p_{\rm dep})\rho,
    \label{eq:depch}
\end{equation}
where $\rho$ is the density matrix of the system, $D$ is the dimension of the underlying Hilbert space, and with probability $p_{\rm dep}$ the system goes to the maximally mixed state $\mathbb{1}/D$. In a general situation, one would be interested in characterizing the average behaviour of a trace preserving quantum channel $\Phi(\rho)$, where $\rho$ is the density matrix of the system ($D\times D$ dimension), averaged over the whole unitary group of dimension $D$, i.e. $U(D)$~\cite{NielsenPLA2002}. If this average commutes with any other unitary $V$ in the group, one can show that this averaged $\Phi$ is equivalent to the depolarizing channel~\cite{EmersonJOptB2005}. In this way, the depolarizing channel provides a kind of ``null hipothesis'' that one can take into account in a decoherence (or error) model. In the context of randomized circuits, the depolarizing channel is used in order to approximate fairly well the source of errors in the mean value of an observable measured, regardless of the origin of the noise~\cite{ZhuPRA2021}. It is also widely used in simulating error correcting codes~\cite{TroutNJP2018}, which makes it fairly ubiquitous in many contexts and many types of circuits.

In our case, we assume that after each step of the protocol, there is an ``accumulated'' decoherence effect that can be modelled using the depolarizing channel. In order to see the effect of the depolarizing channel after $n$ steps, we proceed as follows. For $n$ steps, we require to initialize all the $2^n$ qubits with the same initial condition $(\phi_0,\theta_0)$. This generates the pure initial state $|\Psi_0\rangle$ of the whole circuit which can be written as $\rho_0$. In each step we apply the unitary $U^{(j)}$, which is in general the tensor product of different $U_\epsilon$'s and depends on the qubits that they entangle (see Fig.~\ref{fig:circ_full}). Then, after one step, one has
\begin{equation}
    \rho_1 = U^{(1)}\rho_0 U^{(1)^\dagger}
\end{equation}
and next an accumulated depolarizing effect
\begin{equation}
    \tilde{\rho}_1 = \Phi_{\rm dep}(\rho_1) = (1-p_1)\rho_1 + \frac{p_1}{D}\mathbb{1}.
\end{equation}
Applying this procedure iteratively, one can prove by induction that after $n$ steps
\begin{equation}
    \tilde{\rho}_n = \prod_{j=1}^n (1-p_j)\rho_n + \left(\sum_{j=1}^{n-1}\left[\prod_{l=j+1}^n (1-p_l) \right]\frac{p_j}{D} + \frac{p_n}{D}\right)\mathbb{1}.
    \label{eq:a1}
\end{equation}
The previous equation is still of the form of Eq.~\ref{eq:depch} and a straightforward derivation shows that, if
\begin{equation}
    p_{\rm dep} = \sum_{j=1}^{n-1}p_j\prod_{l=j+1}^n (1-p_l) + p_n,
\end{equation}
then Eq.~\ref{eq:a1} is again a depolarizing channel but now characterized by a global (step dependent) parameter $\tilde{p}$:
\begin{equation}
    \tilde{\rho}_n = (1-p_{\rm dep})\rho_n + \frac{p_{\rm dep}}{D}\mathbb{1}.
\end{equation}
We note that the presented derivation can also be applied to the case where $U^{(j)}$ is some gate inside the circuit for one step of the protocol. That is, one can also consider that $p_{\rm dep}$ after one step of the protocol takes into account the accumulated depolarizing process in the same step. Both descriptions are equivalent.

Note that $\rho_n$ represents the density matrix of the protocol without a depolarizing channel. This implies that the success probability under the influence of the depolarizing channel is
\begin{equation}
    \tilde{p}_{\rm s}^{(n)} = (1-p_{\rm dep}) p_s^{(n)} + \frac{2p_{\rm dep}}{D}.
\end{equation}
In turn, this implies that $\tilde{p}_{\rm s}^{(n)} \leq p_s^{(n)}$ and, more importantly, $\tilde{p}_{\rm s}^{(n)}$ is still constant for fixed $\tilde{p}$ and cannot account for the extra oscillatory effect observed in Figs.~\ref{fig:res1step}, \ref{fig:res2step}, and \ref{fig:res1ps_outsideeps}. The depolarizing channel can only describe a global shift in the mean behaviour of the experimental success probability.

\section{Amplitude damping channel}
\label{sec:ampdamp}
Another relevant source of errors found in the results of the experiments is the contribution of an amplitude damping effect. At zero temperature, the Kraus operators for such a channel can be written as~\cite{Nielsenbook2010}
\begin{equation}
    E_0 = \begin{pmatrix}
        1 & 0 \\ 0 & \sqrt{1-\gamma}\end{pmatrix},\; E_1 = \begin{pmatrix}
            0 & \sqrt{\gamma} \\ 0 & 0
        \end{pmatrix},
        \label{eq:adop}
\end{equation}
where $\gamma$ is interpreted as probability of decaying to the qubit ground state. This probability can be written as $\gamma = 1-e^{t/T_1}$, where $t$ is the time in which states over the Bloch sphere decay to the north pole representing the ground state $|0\rangle$. Here, $T_1$ is a typical time scale and it is particular for every qubit in a given quantum computer. $T_1$ is one of the parameters that describes a qubit in the superconducting IBM quantum computers.

We can compute the modified success probability for one step and one qubit, as we did in the previous case of the depolarizing channel. This modified probability is
\begin{equation}
\tilde{p}^{\rm ad}_s = \frac{1}{(1 + |z|^2)^2}(\varepsilon^2 + |z|^4 + 2\gamma |z|^2 + \gamma(1-\varepsilon^2)),
\end{equation}
where the first two terms represent the error-free success probability $p_s$. In this calculation, we are assuming that after the state preparation and the $U_\epsilon$ gates have been carried out, there is 
an accumulated amplitude damping effect during the circuit execution described by the Kraus operators~\ref{eq:adop}. One can note that the last two terms are always positive, which means that $p_s\leq \tilde{p}^{\rm ad}_s$, i.e. the success probability tends to increment whenever the amplitude damping channel dominates compared to other decoherent processes (for instance, the previous depolarizing channel).

The amplitude damping channel is dominant only in the case of \latdev{nairobi}, in the data shown in Fig.~\ref{fig:res1step}. This is the only quantum computer where the mean of the experimentally estimated success probabilities are shifted above the theoretical one. Indeed, the fitting shown in the figure shows the contribution of the amplitude damping channel combined with a coherent gate error model (see Sec.~\ref{sec:cohmod}), and one can see that the fitting is pretty accurate in describing the displaced, oscillating behaviour of the data. 
We also point out that we only apply the amplitude damping channel to the measured qubit from which the post-selection is made, as this is the only option where one can obtain more counts in $p_s$ (the post-selected qubit can also suffer from amplitude damping, but this is irrelevant from the point of view of $p_s$).

\section{CNOT and the cross resonance-gate}
\label{app:cnot}

In this section we write the single-qubit gates of Eq.~\ref{eq:CNOTquiv} in terms of gates that belong to the \latdev{ibm} operation glossary~\cite{ibmopgloss}. First, regarding the single-qubit gates
\begin{equation}
\begin{split}
    (ZI)^{-1/2} &= {\rm p}_1(-\pi/2) \\
    (IX)^{-1/2} &= {\rm sxdg}_2,\\
\end{split}
\end{equation}
where ${\rm p}_1$ is the phase gate applied to the first qubit and ${\rm sxdg} = \sqrt{X}^\dagger$ is applied to the second qubit. Next, regarding the two-qubit rotation
\begin{equation}
\begin{split}
    U(\beta) &= \exp(-i\beta\pi Z_1 X_2/2)\\ &= \cos\!\left(\!\frac{\beta\pi}{2}\!\right)\mathbb{1}\mathbb{1} - i \sin\!\left(\!\frac{\beta\pi}{2}\!\right)Z_1X_2\\
    &=\!\!\begin{pmatrix}
     \cos\!\left(\!\frac{\beta\pi}{2}\!\right) & -i \sin\!\left(\!\frac{\beta\pi}{2}\!\right) & 0 & 0\\
     -i \sin\!\left(\!\frac{\beta\pi}{2}\!\right) &  \cos\!\left(\!\frac{\beta\pi}{2}\!\right) & 0 & 0 \\
     0 & 0 &  \cos\!\left(\!\frac{\beta\pi}{2}\!\right) & i \sin\!\left(\!\frac{\beta\pi}{2}\!\right) \\
     0 & 0 & i \sin\!\left(\!\frac{\beta\pi}{2}\!\right) &  \cos\!\left(\!\frac{\beta\pi}{2}\!\right)
    \end{pmatrix}
    \end{split} 
\end{equation}
where $\mathbb{1}$ denotes the identity matrix in the two-qubit space. In this case
\begin{equation}
    U(\beta) = J {\rm Cu}(-\beta) J {\rm Cu}(\beta)
\end{equation}
where
\begin{equation}
\begin{split}
     {\rm Cu(\beta)} &= {\rm Cu3}(\beta, \pi/2, -\pi/2)\\
     &= \begin{pmatrix}
         1 & 0 & 0 & 0\\
         0 & 1 & 0 & 0\\
         0 & 0 & \cos(\beta/2) & i\sin(\beta/2) \\
         0 & 0 & i\sin(\beta/2) & \cos(\beta/2)
     \end{pmatrix}
\end{split}
\end{equation}
is the standard qiskit's CU controlled-rotation and $J_{ij} = \delta_{i,N-j+1}$ yields the elements of the exchange matrix $J$~\cite{CantoniLinAlgApp1976}. Both $\beta=1/2$ and the over-rotation are taken into account with the above decomposition.

\section{Gate fidelities and misrotations}
IBM reports the errors of the native gates for every device on its webpage~\cite{qpuinfo_ibm}. Since these data are usually obtained by randomized benchmarking, they describe the performance of the given gate containing many possible sources of errors, therefore, they can be related to the average gate fidelity 
\begin{equation}
    F_{\rm ave} = \frac{D\times F(\mathcal{E},U) + 1}{D+1},
    \label{eq:Fav}
\end{equation}
to the ideal operation $U$. Here, the $F(\mathcal{E},U)$ process fidelity is defined as
\begin{equation}
    F(\mathcal{E},U) = \frac{{\rm Tr}(S_u^\dagger S_\mathcal{E})}{D^2}.
    \label{eq:fid}
\end{equation}
Here $U$ is the expected unitary channel, $\mathcal{E}$ is the actual gate (channel) implemented in the experiments and $S_j$ is the superoperator form of the channel $j$. Equation~\ref{eq:Fav} is obtained in~\cite{horodeckiPRA1999} (see also~\cite{NielsenPLA2002}). Both $F_{\rm ave}$ and $F(\mathcal{E},U)$ are concave, symmetric functions around the angle parameter equal to zero for gates $R_x$ or CR. This symmetry indicates that a given angle error $\alpha$ ($\Lambda$) have the same fidelity as its symmetric counterpart $-\alpha$ ($-\Lambda$). In this way, either $\alpha$ or $\Lambda$ can be considered without loss of generality as positive quantities, albeit in finding the actual values via a numerical computation we still keep potentially signed quantities. 

The error $\epsilon$ in a certain gate defines the fidelity $F_{\rm ave}=1-\epsilon$. One can take this as a reference value in order to estimate the angle $\alpha$ or $\Lambda$ following the process described below. In order to do this, we recovered the corresponding errors $\epsilon$ for the qubits that actually were used to run the circuit. In this way, we compute an average fidelity error $\bar{\epsilon}$ and take $\bar{F}_{\rm av}=1-\bar{\epsilon}$ as the averaged fidelity for the whole set of experiments executed. One can also compute the associated standard deviation of the fidelity $\sigma_F$ and the expression
\begin{equation}
    f = \bar{F}_{\rm av} \pm \sigma_F
\end{equation}
in turn defines an interval in which one can estimate the error in the misrotation $\alpha$ (equivalently $\Lambda$). This ``maximum error angle'' $\alpha^{\rm tol}$ ($\Lambda^{\rm tol}$) is computed numerically using scipy's function \latdev{fsolve} from Eq.~\ref{eq:fid}. This value $\alpha^{\rm tol}$ ($\Lambda^{\rm tol}$) serves as a maximum bound in which one can find the best fitting angles $\alpha$ ($\Lambda$) to a given set of experiments taken from Fig.~\ref{fig:res1step} or~\ref{fig:res2step}.

\section{Computation of the angles of an SU(2) rotation after the transpilation}
\label{app:angles}
In this appendix, we show how to compute the ``Euler angles'' $x,y,z$ for the circuit of Fig.~\ref{fig:circtranspiled}. Recall that the circuit represents the unitary
\begin{equation}
    U = U_\varepsilon U_{\rm st}(\theta_0, \phi_0),
    \label{eq:Ugen}
\end{equation}
where $U_\varepsilon$ is given in Eq.~\ref{eq:Ueps} and $U_{\rm st}$ denotes the state preparation for two qubits, which one can write generically as
\begin{equation}
    U_{\rm st} = U_{\rm st, q1} \otimes  U_{\rm st, q2}=(P(\phi_0)R_y(\theta_0))\otimes (P(\phi_0)R_y(\theta_0)),
    \label{eq:Ust}
\end{equation}
$U_{\rm st, qi}$ denoting the state preparation of qubit $i$, $\phi_0$, $\theta_0$ being the initial angles, $P$ being the phase gate and $R_y$ being the qubit rotation around the $y$-axis. 

\begin{figure}[htb]
  \centering
  \includegraphics[width=0.95\columnwidth]{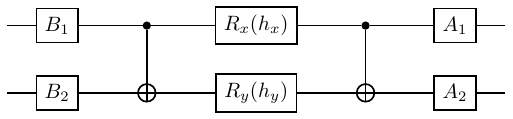}
  \caption{Circuit decomposition of $U_\varepsilon$, as presented in~\cite{OrtegaPLA2021}.}
  \label{fig:Uepscirc}
\end{figure}
The unitary $U$ has to be transpiled into the set of native gates of the given quantum computer. In the tested quantum computers, this set is composed of $\sqrt{X}$, $R_z$ and CNOT gates. Note that, following the previous deocomposition found in~\cite{OrtegaPS2023}, $U_\varepsilon$ can be decomposed into a series of one-qubit gates and two CNOTs. Generically, the resulting unitary is shown in Fig.~\ref{fig:Uepscirc} for $A_1,A_2,B_1,B_2$ general SU(2) rotations and $R_x,R_y$ qubit rotations over the axis $x$ and $y$ respectively (the angles $h_x$ and $h_y$ can be computed as described in~\cite{OrtegaPS2023}). In the full $U$ matrix, which includes the multiplication of $U_{\rm st}$ and $U_\varepsilon$, before the first CNOT in $U_\varepsilon$, we have a multiplication of two one-qubit SU(2) matrices, those coming from the state preparation in Eq.~\ref{eq:Ust} and those coming from the $U_\varepsilon$ decomposition. Since we fix $\varepsilon$ (and $\theta_0$), the only parameter that changes in the unitary is $\phi_0$, which in turn means that the product of SU(2) matrices before the first CNOT, mentioned above, is the only one that depends on $\phi_0$. Yet, note that the circuit in Fig.~\ref{fig:circtranspiled} only contain as variables the angles $x,y,z$, before the application of the first CNOT. In general, $x,y,z$ depend explicitly on $\phi_0, \theta_0, \varepsilon$, as described below. Thus, one needs to compute such angles in order to write a general SU(2) as a combination of the allowed gates $\sqrt{X}$ and $R_x$.

It is possible to write a matrix that resembles an SU(2) using the gate combinations of $\sqrt{X}, R_z$ as
\begin{multline}
    R_z(x)\sqrt{X} R_z(y) \sqrt{X}R_z(w) = \\
    =\begin{pmatrix}
        e^{-i(x+w)/2}\sin\left(\frac{y}{2}\right) & e^{-i(x-w)/2}\cos\left(\frac{y}{2}\right) \\
        e^{i(x-w)/2}\cos\left(\frac{y}{2}\right) & -e^{i(x+w)/2}\sin\left(\frac{y}{2}\right)
    \end{pmatrix}.
\end{multline}
But now it is readily seen that the above matrix is in U(2) and not SU(2) as in Eq.~\ref{eq:Ugen}, but one can make the above a member of SU(2) by multiplying it by the global phase $e^{i\pi/2}$. Thus, for qubit $i$,
\begin{equation}
    R_z(x_i)\sqrt{X} R_z(y_i) \sqrt{X}R_z(w_i) = e^{i\pi/2}B_i \times U_{\rm st, qi}.
    \label{eq:xyz}
\end{equation}
Using this equality, one can find by direct substitution the values of the angles $x,y,z$. If $A =e^{i\pi/2}B_i \times U_{\rm st, q1}$ and the elements of $A$ are denoted by $a_{ij}$, the values of the angles read
\begin{equation}
    \begin{split}
        y &= 2 \cos^{-1}(\sqrt{a_{12}a_{21}}),\\
        x &= i\log\left(\frac{a_{11}}{a_{21}\tan(y/2)} \right),\\
        w &= -i \log\left(-\frac{a_{22}}{a_{21}\tan(y/2)} \right).
    \end{split}
\end{equation}

\section{Fitting parameters of the coherent error model}
\begin{table}[h]
\resizebox{\columnwidth}{!}{%
    \begin{tabular} {|c|c|c|c|c|c|}
        \hline
        Device  & $\alpha_1$ $(\alpha_1^{\rm tol})$ & $\alpha_2$ $(\alpha_2^{\rm tol})$ & $\Lambda$ $(\Lambda^{\rm tol})$ & $p_{\rm dep}$ & $\gamma$ \\ \hline
        nairobi & $2.83$ $(\pm 5.26)$ & $4.45$ $(\pm 4.45)$ & $-1.78$ $(\pm 7.25)$  & - & $1.21$ \\ 
        lima & $3.63$ $(\pm 7.55)$ & $4.83$ $(\pm 5.73)$ & $-2.63$ $(\pm 6.68)$ & $1.99$ & - \\ 
        manila & $-0.55$ $(\pm 4.38)$ & $0.29$ $(\pm 3.88)$ & $0.06$ $(\rm 6.43)$ & $2.30$ & - \\
        quito & $-4.30$ $(\pm 4.30)$ & $1.22$ $(\pm 5.48)$ & $1.26$ $(\pm 8.13)$ & $2.64$ & - \\
        oslo & $3.44$ $(\pm 3.44)$ & $0.61$ $(\pm 7.85)$ & $-2.31$ $(\pm 7.63)$ & $6.46$ & - \\
        \hline
    \end{tabular}} \\
    \caption{Fitted parameters (in units of $10^{-2}$) corresponding to the test with one iteration of the protocol, for the fits shown in Fig.~\ref{fig:res1step}.}
    \label{tab:params_1step}   
\end{table}
\begin{table}[h]
\resizebox{\columnwidth}{!}{%
    \begin{tabular} {|c|c|c|c|c|}
        \hline
        Device & $\alpha_1$ $(\alpha_{1}^{\rm tol})$ &  $\alpha_2$ $(\alpha_2^{\rm tol})$ & $\alpha_3$ $(\alpha_3^{\rm tol})$ & $\alpha_4$ $(\alpha_4^{\rm tol})$ \\ \hline
        nairobi & $2.74$  $(\pm 4.04)$ & $-2.98$ $(\pm 4.33)$ & $5.04$ $(\pm 5.04)$ & $4.05$ $(\pm 4.05)$ \\
        lima & $3.58$ $(\pm 3.58)$ & $2.11$ $(\pm 3.25)$ & $3.99$ $(\pm 3.99)$ & $-1.86$ $(\pm 6.29)$ \\
        bogota & $4.47$ $(\pm 4.47)$ & $0.83$ $(\pm 4.05)$ & $3.24$ $(\pm 3.24)$ & $5.27$ $(\pm 6.56)$ \\ 
        manila & $4.38$ $(\pm 4.38)$ & $2.34$ $(\pm 3.88)$ & $-5.05$ $(\pm 5.05)$ & $3.56$ $(\pm 3.56)$ \\ 
        quito & $-4.47$ $(\pm 4.47)$ & $2.75$ $(\pm 4.68)$ & $5.00$ $(\pm 5.00)$ & $5.23$ $(\pm 5.23)$ \\
        santiago & $7.71$ $(\pm 7.71)$ & $-3.18$ $(\pm 3.18)$ & $-3.52$ $(\pm 3.52)$ & $4.58$ $(\pm 4.58)$ \\
        oslo & $0.89$ $(\pm 4.34)$ & $4.59$ $(\pm 4.59)$ & $3.38$ $(\pm 5.93)$ & $4.92$ $(\pm 4.92)$ \\
        \hline
    \end{tabular}} \\
    \vspace{1ex}
    \resizebox{\columnwidth}{!}{%
     \begin{tabular} {|c|c|c|c|c|c|}
        \hline
        Device & $\Lambda_1$ $(\Lambda_1^{\rm tol})$ & $\Lambda_2$ $(\Lambda_2^{\rm tol})$ & $\Lambda_3$ $(\Lambda_3^{\rm tol})$ & $\Lambda_4$ $(\Lambda_4^{\rm tol})$ & $p_{\rm dep}$ \\ \hline
        nairobi & $0.73$ $(\pm 7.22)$ & $3.78$ $(\pm 9.62)$ & $-0.85$ $(\pm 7.28)$ & $7.28$ $(\pm 7.28)$ & $6.42$ \\
        lima & $-4.42$ $(\pm 5.05)$ & $3.13$ $(\pm 9.53)$ & $8.44$ $(\pm 8.44)$ & $3.33$ $(\pm 8.44)$ & $10.48$ \\
        bogota & $-7.71$ $(\pm 7.71)$ & $2.00$ $(\pm 6.84)$ & $1.27$ $(\pm 6.29)$ & $1.15$ $(\pm 6.29)$ & $12.77$ \\ 
        manila & $-4.50$ $(\pm 6.43)$ & $8.32$ $(\pm 30.18)$ & $-0.97$ $(\pm 30.18)$ & $4.89$ $(\pm 8.00)$ & $14.60$ \\ 
        quito & $6.29$ $(\pm 6.40)$ & $-2.08$ $(\pm 9.87)$ & $-0.17$ $(\pm 9.87)$ & $-7.54$ $(\pm 7.71)$ & $15.10$ \\
        santiago & $-1.50$ $(\pm 10.71)$ & $9.27$ $(\pm 13.47)$ & $6.14$ $(\pm 13.47)$ & $4.79$ $(\pm 11.04)$ & $16.90$ \\
        oslo & $-5.90$ $(\pm 7.47)$ & $-1.47$ $(\pm 6.17)$ & $6.17$ $(\pm 6.17)$ & $4.79$ $(\pm 7.44)$ & $23.21$ \\
        \hline
    \end{tabular}} \\
    \caption{Fitted parameters (in units of $10^{-2}$) corresponding to the test with two iterations of the protocol, for the fits shown in Fig.~\ref{fig:res2step}.}
    \label{tab:params_2step}   
\end{table}

\section{Dates of the experiments}
\label{sec:app_dates}

In this appendix we provide the dates of the experiments run in the  devices. 
\begin{table}[h]
    \begin{tabular} {|c|c|}
        \hline
        Device & Date(s)\\ \hline
        nairobi & 9-15 Nov 2022 \\ 
        lima & 3-8 Nov 2022 \\ 
        manila & 6 June -15 Aug 2023 \\
        quito & 3-12 Nov 2022\\ 
        oslo & 9-15 Nov 2022\\
        \hline
    \end{tabular} \\
    \vskip .5cm
    \caption{Dates for the experiments presented in Fig.~\ref{fig:res1step}.}
    \label{tab:dates1}   
\end{table}
\begin{table}[h]
    \begin{tabular} {|c|c|}
        \hline
        Device & Date(s)\\ \hline
        nairobi &6-9 July 2022\\ 
        lima & 17-22 May 2022\\ 
        bogota & 17-20 May 2022\\
        manila & 2-14 Aug 2023\\
        quito & 24 May - 2 June 2022\\ 
        santiago & 24-27 May 2022\\
        oslo & 6-9 July 2022\\
        \hline
    \end{tabular} 
    \caption{Dates for the experiments presented in Fig.~\ref{fig:res2step}.}
    \label{tab:dates2}   
\end{table}
\begin{table}[h]
\resizebox{\columnwidth}{!}{%
    \begin{tabular} {|c|c|c|c|}
        \hline
        qubits & lima & quito & nairobi \\ \hline
        (01) & 21-24 Nov 2022 & 21 Nov - 5 Dec 2022 & 25 Nov - 6 Dec 2022 \\ 
        (12) & 10-11 Jan 2023 & 20 Feb - 10 Mar 2023 & 9-18 Aug 2023 \\ 
        (13) & 11 Jan 2023 & 13-14 Mar 2024 & 20 Jan - 3 Feb 2023 \\
        (34) & 11-12 Jan 2023 & 14 Mar - 3 Apr 2023 & - \\
        (35) & -& -& 6-14 Feb 2023 \\   
        (45) & -& -& 26 July - 9 Aug 2023 \\
        (56) & -& -& 18-22 Aug 2023 \\
        \hline
    \end{tabular}} \\
    \caption{Dates for the experiments presented in  Fig.~\ref{fig:res1ps_outsideeps}.}
    \label{tab:dates3}   
\end{table}
\begin{table}[h]
\resizebox{\columnwidth}{!}{%
    \begin{tabular} {|c|c|c|c|}
        \hline
        qubits & lima & quito & nairobi \\ \hline
        (0134) & 12-19 Dec 2022 & 12 Dec 2022 & - \\ 
        (2134) & 6-9 Feb 2023 & 13 Apr 2023 & - \\ 
        (0135) & - & - & 13 Dec 2022 - 3 Jan 2023\\ 
        (2135) & - & - & 14 Aug - 14 Sep 2023\\ 
        \hline
    \end{tabular}} \\
    \caption{Dates for the experiments presented in  Fig.~\ref{fig:res1ps_outsideeps_twosteps}.}
    \label{tab:dates4}   
\end{table}
\bibliographystyle{hunsrt}
\bibliography{bibliography.bib}

\begin{thebibliography}{10}

\bibitem{Vieiraext2013}
M.~Vieira, H.~Madeira, K.~Sachs, and S.~Kounev.
\newblock Resilience benchmarking.
\newblock page 283. Springer, 1 edition, 2012.

\bibitem{Liljabook2000}
D.~J. Lilja.
\newblock {\em Measuring computer performance}.
\newblock Cambridge University Press, 1 edition, 2000.

\bibitem{Kounevbook2020}
S.~Kounev, K-D Lange, and J~von Kistowski.
\newblock {\em Systems Benchmarking}.
\newblock Springer, 1 edition, 2020.

\bibitem{ReschACM2021}
Salonik Resch and Ulya~R. Karpuzcu.
\newblock Benchmarking quantum computers and the impact of quantum noise.
\newblock {\em {ACM} {C}omputing {S}urveys}, 54:142, 2021.

\bibitem{ProctorNatPhys2021}
T.~Proctor, K.~Rudinger, K.~Young, E.~Nielsen, and R.~Blume-Kohout.
\newblock Measuring the capabilities of quantum computers.
\newblock {\em Nat. Phys.}, 18:75, 2021.

\bibitem{BlumeOSTI2019}
Robin~J Blume-Kohout and Kevin Young.
\newblock Metrics and benchmarks for quantum processors: State of play.
\newblock 1 2019.

\bibitem{AcuavivaArxiv2024}
Arturo Acuaviva, David Aguirre, and Mikel~Sanz Rub\'{e}n Pe\~{n}a.
\newblock Benchmarking quantum computers: Towards a standard performance
  evaluation approach, 2024, arXiv:2407.10941.

\bibitem{ProctorArxiv2024}
Timothy Proctor, Kevin Young, Andrew~D. Baczewski, and Robin Blume-Kohout.
\newblock Benchmarking quantum computers, 2024, arXiv:2407.08828.

\bibitem{KnillPRA2008}
E.~Knill, D.~Leibfried, R.~Reichle, J.~Britton, R.~B. Blakestad, J.~D. Jost,
  C.~Langer, R.~Ozeri, S.~Seidelin, and D.~J. Wineland.
\newblock Randomized benchmarking of quantum gates.
\newblock {\em Phys. Rev. A}, 77:012307, 2008.

\bibitem{MagesanPRL2011}
Easwar Magesan, J.~M. Gambetta, and Joseph Emerson.
\newblock Scalable and robust randomized benchmarking of quantum processes.
\newblock {\em Phys. Rev. Lett.}, 106:180504, 2011.

\bibitem{MagesanPRA2012}
Easwar Magesan, Jay~M. Gambetta, and Joseph Emerson.
\newblock Characterizing quantum gates via randomized benchmarking.
\newblock {\em Phys. Rev. A}, 85:042311, 2012.

\bibitem{ProctorPRL2019}
Timothy~J. Proctor, Arnaud Carignan-Dugas, Kenneth Rudinger, Erik Nielsen,
  Robin Blume-Kohout, and Kevin Young.
\newblock Direct randomized benchmarking for multiqubit devices.
\newblock {\em Phys. Rev. Lett.}, 123:030503, 2019.

\bibitem{CrossPRA2019}
Andrew~W. Cross, Lev~S. Bishop, Sarah Sheldon, Paul~D. Nation, and Jay~M.
  Gambetta.
\newblock Validating quantum computers using randomized model circuits.
\newblock {\em Phys. Rev. A}, 100:032328, Sep 2019.

\bibitem{ProctorPRL2022}
Timothy Proctor, Stefan Seritan, Kenneth Rudinger, Erik Nielsen, Robin
  Blume-Kohout, and Kevin Young.
\newblock Scalable randomized benchmarking of quantum computers using mirror
  circuits.
\newblock {\em Phys. Rev. Lett.}, 129:150502, 2022.

\bibitem{ChenPRA2023}
Jianxin Chen, Dawei Ding, Cupjin Huang, and Linghang Kong.
\newblock Linear cross-entropy benchmarking with clifford circuits.
\newblock {\em Phys. Rev. A}, 108:052613, 2023.

\bibitem{MesmanArxiv2021}
Koen Mesman, Zaid Al-Ars, and Matthias M\"{o}ller.
\newblock Qpack: Quantum approximate optimization algorithms as universal
  benchmark for quantum computers, 2021, arXiv:2103.17193.

\bibitem{LubinskiArxiv2023}
Thomas Lubinski, Carleton Coffrin, Catherine McGeoch, Pratik Sathe, Joshua
  Apanavicius, and David E.~Bernal Neira.
\newblock Optimization applications as quantum performance benchmarks, 2023,
  arXiv:2302.02278.

\bibitem{LinkePNAS2017}
N.~M. Linke, D~Maslovc, M.~Roettelerd, S.~Debnath, C.~Figgatt, K.~A. Landsman,
  K.~Wright, and C.~Monroe.
\newblock Experimental comparison of two quantumcomputing architectures.
\newblock {\em PNAS}, 114:3305, 2017.

\bibitem{WrightNatCom2019}
K.~Wright, K.~M. Beck, S.~Debnath, J.~M. Amini, Y.~Nam, N.~Grzesiak, J.-S.
  Chen, N.~C. Pisenti, M.~Chmielewski, C.~Collins, K.~M. Hudek, J.~Mizrahi,
  J.~D. Wong-Campos, S.~Allen, J.~Apisdorf, P.~Solomon, M.~Williams, A.~M.
  Ducore, A.~Blinov, S.~M. Kreikemeier, V.~Chaplin, M.~Keesan, C.~Monroe, and
  J.~Kim.
\newblock Benchmarking an 11-qubit quantum computer.
\newblock {\em Nature {C}ommunications}, 10:5464, 2019.

\bibitem{GilyenArxiv2021}
Arjan Cornelissen, Johannes Bausch, and András Gilyén.
\newblock Scalable benchmarks for gate-based quantum computers, 2021,
  arXiv:2104.10698.

\bibitem{McCaskeyNQI2019}
A.~J. McCaskey, Z.~P. Parks, J.~Jakowski, S.~V. Moore, T.~D. Morris, T.~Humble,
  and R.~C. Pooser.
\newblock Quantum chemistry as a benchmark for near-term quantum computers.
\newblock {\em npj Quantum Inf.}, 5:99, 2019.

\bibitem{FinzgarArxiv2022}
Jernej~Rudi Fin\v{z}gar, Philipp Ross, Leonhard Hölscher, Johannes Klepsch,
  and Andre Luckow.
\newblock Quark: A framework for quantum computing application benchmarking,
  2022, arXiv:2202.03028.

\bibitem{TomeshArxiv2022}
Teague Tomesh, Pranav Gokhale, Victory Omole, Gokul~Subramanian Ravi,
  Kaitlin~N. Smith, Joshua Viszlai, Xin-Chuan Wu, Nikos Hardavellas,
  Margaret~R. Martonosi, and Frederic~T. Chong.
\newblock Supermarq: A scalable quantum benchmark suite, 2022,
  arXiv:2202.11045.

\bibitem{LiArxiv2020}
Ang Li, Samuel Stein, Sriram Krishnamoorthy, and James Ang.
\newblock Qasmbench: A low-level qasm benchmark suite for nisq evaluation and
  simulation, 2020, arXiv:2005.13018.

\bibitem{DasguptaArxiv2020}
Samudra Dasgupta and Travis~S. Humble.
\newblock Characterizing the stability of {NISQ} devices, 2020,
  arXiv:quant-ph/2008.09612.

\bibitem{KissPRA2006}
T.~Kiss, I.~Jex, G.~Alber, and S.~Vym\v{e}tal.
\newblock Complex chaos in the conditional dynamics of qubits.
\newblock {\em Phys. Rev. A}, 74:040301(R), 2006.

\bibitem{GilyenSciRep2016}
András Gily\'{e}n, Tam\'{a}s Kiss, and Igor Jex.
\newblock Exponential sensitivity and its cost in quantum physics.
\newblock {\em Sci. Rep.}, 6:20076, 2016.

\bibitem{MalachovChaos2019}
Martin Malachov, Igor Jex, Orsolya K\'{a}lm\'{a}n, and Tam\'{a}s Kiss.
\newblock Phase transition in iterated quantum protocols for noisy inputs.
\newblock {\em Chaos}, 29:033107, 2019.

\bibitem{ViennotCSF2022}
David Viennot.
\newblock Competition between decoherence and purification: Quaternionic
  representation and quaternionic fractals.
\newblock {\em Chaos Solit.}, 161:112346, 2022.

\bibitem{GuanPRA2013}
Yilun Guan, Duy~Quang Nguyen, Jingwei Xu, and Jiangbin Gong.
\newblock Reexamination of measurement-induced chaos in
  entanglement-purification protocols.
\newblock {\em Phys. Rev. A}, 87:052316, 2013.

\bibitem{PortikPLA2022}
Attila Portik, Orsolya K\'{a}lm\'{a}n, Igor Jex, and Tam\'{a}s Kiss.
\newblock Iterated nth order nonlinear quantum dynamics with mixed initial
  states.
\newblock {\em Phys. Lett. A}, 431:127999, 2022.

\bibitem{PortikPRA2024}
Attila Portik, Orsolya K\'{a}lm\'{a}n, Igor Jex, and Tam\'{a}s Kiss.
\newblock Robustness of chaotic behavior in iterated quantum protocols.
\newblock {\em Phys. Rev. A}, 109:042410, 2024.

\bibitem{TorresPRA2017}
Juan~Mauricio Torres, J\'{o}zsef~Zsolt Bern\'{a}d, Gernot Alber, Orsolya
  K\'{a}lm\'{a}n, and Tam\'{a}s Kiss.
\newblock Measurement-induced chaos and quantum state discrimination in an
  iterated tavis-cummings scheme.
\newblock {\em Phys. Rev. A}, 95:023828, 2017.

\bibitem{KalmanPRA2018}
Orsolya K\'alm\'an and Tam\'as Kiss.
\newblock Quantum state matching of qubits via measurement-induced nonlinear
  transformations.
\newblock {\em Phys. Rev. A}, 97:032125, Mar 2018.

\bibitem{OrtegaPS2023}
A.~Ortega, O.~K\'{a}lm\'{a}n, and T.~Kiss.
\newblock Testing quantum computers with the protocol of quantum state
  matching.
\newblock {\em Phys. Scr.}, 98:024006, 2023.

\bibitem{TucciArxiv2005}
Robert~R. Tucci.
\newblock An introduction to {C}artan's {KAK} decomposition for {QC}
  programmers, 2005, arXiv:quant-ph/0507171.

\bibitem{VidalPRA2004}
G.~Vidal and C.~M. Dawson.
\newblock Universal quantum circuit for two-qubit transformations with three
  controlled-not gates.
\newblock {\em Phys. Rev. A}, 69:010301, Jan 2004.

\bibitem{ParisLNP2004}
M~Paris and J~\v{R}eh\'{a}\v{c}ek, editors.
\newblock {\em Quantum State Estimation}, volume 649 of {\em Lecture Notes in
  Physics}. Springer, 2004.

\bibitem{IBMqcs}
{IBM Quantum}.
\newblock \url{ https://quantum-computing.ibm.com/}.
\newblock Accessed: 2022-06.

\bibitem{ZimborasQuantum2022}
F~B Maciejewski, Z~Zimbor\'as, and M~Oszmaniec.
\newblock Mitigation of readout noise in near-term quantum devices by classical
  post-processing based on detector tomography.
\newblock {\em Quantum}, 6:707, 2022.

\bibitem{MakhlinRMP2001}
Yuriy Makhlin, Gerd Sch\"on, and Alexander Shnirman.
\newblock Quantum-state engineering with josephson-junction devices.
\newblock {\em Rev. Mod. Phys.}, 73:357--400, May 2001.

\bibitem{ClarkeNat2008}
J.~Clarke and F.~K. Wilhem.
\newblock Superconducting quantum bits.
\newblock {\em Nature}, 453:1031, 2008.

\bibitem{KrantzAPR2019}
P.~Krantz, M.~Kjaergaard, F.~Yan, T.~P. Orlando, S~Gustavsson, and W.~D.
  Oliver.
\newblock A quantum engineer's guide to superconducting qubits.
\newblock {\em Appl. Phys. Rev.}, 6:021318, 2019.

\bibitem{ShorPRA1995}
Peter~W. Shor.
\newblock Scheme for reducing decoherence in quantum computer memory.
\newblock {\em Phys. Rev. A}, 52:R2493--R2496, Oct 1995.

\bibitem{AruteNat2019}
Babbush R. et~al. Arute~F., Arya~K.
\newblock Quantum supremacy using a programmable superconducting processor.
\newblock {\em Nature}, 574:505, 2019.

\bibitem{TroutNJP2018}
C.~J. Trout, M.~Li, M.~Guti\'errez, Y.~Wu, S-T Wang, L.~Duan, and K.~R Brown.
\newblock Simulating the performance of a distance-3 surface code in a linear
  ion trap.
\newblock {\em New J. Phys.}, 20:043038, 2018.

\bibitem{ZhuPRA2021}
D.~Zhu, S.~Johri, N.~H. Nguyen, C.~Huerta Alderete, K.~A. Landsman, N.~M.
  Linke, C.~Monroe, and A.~Y. Matsuura.
\newblock Probing many-body localization on a noisy quantum computer.
\newblock {\em Phys. Rev. A}, 103:032606, Mar 2021.

\bibitem{mckayPRA2017}
David~C. McKay, Christopher~J. Wood, Sarah Sheldon, Jerry~M. Chow, and Jay~M.
  Gambetta.
\newblock Efficient $z$ gates for quantum computing.
\newblock {\em Phys. Rev. A}, 96:022330, Aug 2017.

\bibitem{RigettiPRB2010}
Chad Rigetti and Michel Devoret.
\newblock Fully microwave-tunable universal gates in superconducting qubits
  with linear couplings and fixed transition frequencies.
\newblock {\em Phys. Rev. B}, 81:134507, Apr 2010.

\bibitem{AlexanderQST2020}
T.~Alexander, N.~Kanazawa, D.~J. Egger, L.~Capelluto, C.~J. Wood,
  A.~Javadi-Abhari, and D.~C. McKay1.
\newblock Qiskit pulse: programming quantum computers through the cloud with
  pulses.
\newblock {\em Quantum Sci. Technol.}, 5:044006, 2020.

\bibitem{SmithCommACM1988}
J.~E. Smith.
\newblock Characterizing computer performance with a single number.
\newblock {\em Comm. of the {ACM}}, 31:1202, 1988.

\bibitem{NielsenPLA2002}
M.~A. Nielsen.
\newblock A simple formula for the average gate fidelity of a quantum dynamical
  operation.
\newblock {\em Phys. Lett. A}, 303:249, 2002.

\bibitem{EmersonJOptB2005}
Joseph Emerson, Robert Alicki, and Karol Życzkowski.
\newblock Scalable noise estimation with random unitary operators.
\newblock {\em Journal of Optics B: Quantum and Semiclassical Optics},
  7(10):S347, 2005.

\bibitem{Nielsenbook2010}
M.~A. Nielsen and I.~L. Chuang.
\newblock {\em Quantum Computation and Quantum Information}.
\newblock Cambridge University Press, 10th ann. ed. edition, 2010.

\bibitem{ibmopgloss}
{IBM} operation glossary.
\newblock
  \url{https://quantum-computing.ibm.com/composer/docs/iqx/operations_glossary/}.
\newblock Accessed: 2023-10.

\bibitem{CantoniLinAlgApp1976}
A.~Cantoni and P.~Butler.
\newblock Eigenvalues and eigenvectors of symmetric centrosymmetric matrices.
\newblock {\em Linear Algebra and its Applications}, 13(3):275--288, 1976.

\bibitem{qpuinfo_ibm}
{QPU information}.
\newblock \url{https://docs.quantum.ibm.com/guides/qpu-information}.
\newblock Accessed: 2024-08.

\bibitem{horodeckiPRA1999}
Micha\l{} Horodecki, Pawe\l{} Horodecki, and Ryszard Horodecki.
\newblock General teleportation channel, singlet fraction, and
  quasidistillation.
\newblock {\em Phys. Rev. A}, 60:1888--1898, Sep 1999.

\bibitem{OrtegaPLA2021}
A~Ortega, T~Gorin, and C~S Hamilton.
\newblock Quantum transport in a combined kicked rotor and quantum walk system.
\newblock {\em Phys. Lett. A}, 395:127224, 2021.

\end{thebibliography}

\end{document}